\documentclass{aa}  

\usepackage{graphicx}
\usepackage{txfonts}
\usepackage{xcolor}
\usepackage{xspace}
\usepackage{mathrsfs}
\newcommand{\ohhdp}{o$\rm H_2D^+$\xspace}
\newcommand{\hhdp}{$\rm H_2D^+$\xspace}
\newcommand{\xe}{$x(\mathrm{e^-})$\xspace}

\newcommand{\cdo}{$\rm C^{18}O$\xspace}
\newcommand{\htcop}{$\rm H^{13}CO^+$\xspace}
\newcommand{\dcop}{$\rm DCO^+$\xspace}
\newcommand{\hcope}{$\rm HCO^+$\xspace}
\newcommand{\tex}{$T\rm _{ex}$\xspace}
\newcommand{\ncol}{$N\rm _{col}$\xspace}
\newcommand{\tdust}{$T\rm _{dust}$\xspace}

\newcommand{\olineh}{o$\rm H_2D^+(1_{1,0} - 1_{1,1})$\xspace}
\newcommand{\crir}{$\zeta_2$\xspace}

\newcommand{\kms}{$\rm km \, s^{-1}$\xspace}

\begin{document}

   \title{Cosmic-ray ionisation rate in low-mass cores: the role of the environment}

\author{E. Redaelli\inst{1,2} 
\and S. Bovino\inst{3,4,5} \and G. Sabatini\inst{3,6} \and D. Arzoumanian\inst{7, 8, 9} \and M. Padovani\inst{3} \and P. Caselli \inst{2} \and F. Wyrowski\inst{10} \and J. E. Pineda\inst{2} \and G.  Latrille \inst{5}  }

\institute{
European Southern Observatory, Karl-Schwarzschild-Stra{\ss}e 2, 85748 Garching, Germany \and
Max-Planck-Institut f\"ur Extraterrestrische Physik, Giessenbachstrasse 1, 85748 Garching, Germany 
\and  INAF, Osservatorio Astrofisico di Arcetri, Largo E. Fermi 5, I-50125, Firenze, Italy
        \and
        Chemistry Department, Sapienza University of Rome, P.le A. Moro, 00185 Rome, Italy
        \and 
        Departamento de Astronom\'ia, Facultad Ciencias F\'isicas y Matem\'aticas, Universidad de Concepci\'on, Av. Esteban Iturra s/n Barrio Universitario, Casilla 160, Concepci\'on, Chile \and  
        INAF, Istituto di Radioastronomia - Italian node of the ALMA Regional Centre (It-ARC), Via Gobetti 101, I-40129 Bologna, Italy \and
         Institute for Advanced Study, Kyushu University, Japan
\and Department of Earth and Planetary Sciences, Faculty of Science, Kyushu University, Nishi-ku, Fukuoka 819-0395, Japan
\and  National Astronomical Observatory of Japan, 2-21-1 Osawa, Mitaka, Tokyo 181-8588, Japan
  \and
        Max-Planck-Institut f\"ur Radioastronomie, Auf dem H\"ugel, 69, 53121, Bonn, Germany 
       }

   \date{XXX}

 
  \abstract
   {Cosmic rays drive several key processes for the chemistry and dynamical evolution of star-forming regions. Their effect is quantified mainly by means of the cosmic-ray ionisation rate \crir. }
   {{We aim to obtain a sample of \crir measurements in 20} low-mass starless cores embedded in different parental clouds, to assess the average level of ionisation in this kind of sources and to investigate the role of the environment in this context. {The warmest clouds in our sample are Ophiuchus and Corona Australis, where star formation activity is higher than in the Taurus cloud and the other isolated cores we targeted.}}
   {We {compute} \crir using an analytical method based on the {column density} of ortho-$\rm H_2D^+$, the CO abundance, and the deuteration level of \hcope. To estimate these quantities{,} we analysed new, high-sensitivity molecular line observations obtained with the Atacama Pathfinder EXperiment (APEX) single-dish telescope {and archival continuum data from \textit{Herschel}}.}
   {We report \crir estimates in 17 cores in our sample {and provide} upper limits on the three remaining sources. The values  span almost two orders of magnitude, from $1.3 \times 10^{-18}\, \rm s^{-1}$ to $8.5 \times 10^{-17}\, \rm s^{-1}$.}
  {We find no significant correlation between \crir and the core's column densities $N\rm (H_2)$. On the contrary, we find a positive correlation between \crir and the cores' temperature, estimated via \textit{Herschel} data: cores embedded in warmer environments present higher ionisation levels. The warmest clouds in our sample are Ophiuchus and Corona Australis, where star formation activity is higher than in the other clouds we targeted. {The higher ionisation rates in these regions support the scenario that low-mass protostars in the vicinity of our targeted cores contribute to the re-acceleration of local cosmic rays.} }


   \maketitle
%

\section{Introduction}
Cosmic rays (CRs) are energetic, ionised particles found ubiquitously in the interstellar medium (ISM). In the densest gas phases that represent the initial stage of star formation, the high visual extinction {($A_\mathrm{V}\gtrsim 6 \rm \, mag$ is considered the threshold for star formation, cf. \citealt{Froebrich10, Pineda23})} completely attenuates the ultraviolet flux. {In these conditions, without the existence of X-rays, CRs are the sole ionising agent}. CRs ionise H$_2$ molecules and produce H$_3^+$, which in turn determines the ionisation fraction within dense matter. This ionisation fraction plays a crucial role in regulating star formation by controlling the rate of ambipolar diffusion, a process that can slow down gravitational collapse \citep{Mouschovias76}. H$_3^+$, moreover, is a pivotal species for the chemical evolution of star-forming regions, because it drives the efficient ion chemistry. By reacting with deuterated hydrogen HD,  H$_3^+$ is converted into \hhdp, which is the starting point of the deuteration process in the {cold molecular} gas phase \citep[see e.g.][and references therein]{Ceccarelli14}. CRs deeply impact the chemistry and physics of the dense ISM \citep[see][for a review on the topic]{Padovani20}. \par
Estimating the CR ionisation rate (\crir, expressed relative to $\rm H_2$ molecules) from observations is still challenging. In the diffuse ISM ($A_\mathrm{V}\lesssim 1 \, \rm mag$), H$_3^+$ absorption lines {against bright background OB stars can be directly used} to infer \crir. This is the approach followed for instance by \cite{McCall03,Indriolo07, Indriolo12}, with typical results of the order of $\text{\crir} \gtrsim 10^{-16} \, \rm s^{-1}$. This method, however, relies on the estimate of the gas volume density $n \rm (H_2)$. Traditionally, this parameter has been estimated by fitting the populations of excited rotational states of the molecule C$_2$, taking into account infrared pumping and collisional de-excitation \citep{VanDishoeck82}. In a recent work, however, \cite{Obolentseva24} re-evaluated \crir towards 12 diffuse lines of sight, using 3D gas density distributions obtained from extinction maps \citep{Edenhofer24}, which leads to lower $n \rm (H_2)$ values. As a result, the computed \crir values are also lower by almost one order of magnitude. Incidentally, the new density estimates are in good agreement with the values obtained revising the available C$_2$ data \citep{Neufeld24}.
\par
In the {cold dense} gas, a value of $\zeta_2 \approx 1-3 \times 10^{-17} \rm \, s^{-1}$ is often assumed, for instance in chemical modelling or magneto-hydrodynamic simulations \citep[cf.][]{Sipila15, Kong15, Bovino19}. Observational estimates, however, are spread over a wider range, depending also on the methodology used. {At visual extinction higher than a few $\rm mag$, the absorption lines of H$_3^+$ cannot be detected anymore}. {\cite{Guelin77, Guelin82} proposed an analytic expression to infer the ionisation fraction based on the kinetics of \dcop{} and \hcope{}, and \cite{Wootten79} expanded the method to obtain constraints on \crir}.
    \cite{Caselli98} later expanded the research, reporting the first large ($\sim 20$ objects) sample of \crir measurements in low-mass dense cores. Those authors used both an analytical method and a more detailed analysis using a chemical code. {The analytical method, however, has been found to have strong limitations (see \citealt{Caselli98} itself; \citealt{Caselli02b}; more recently, \citealt{Redaelli24})}. Several other works tackled {the problem of measuring \crir at high densities}, yielding to \crir values scattered over a large range \citep{Maret07, Hezareh08,Fuente16, Redaelli21b, Harju24}, typically from $ \sim 10^{-18} \rm \, s^{-1}$ to $ \text{a few} \times 10^{-16} \rm \, s^{-1}$. 
\par
{More recently, \cite{Bovino20} proposed an analytical approach to determine \crir {in dense prestellar cores} from the detection of \hhdp (and, in particular, of its ortho spin state, \ohhdp, which can be observed from the ground), starting from the formulation of \cite{Oka19}. It is based on inferring the H$_3^+$ abundance from that of its deuterated forms, using the deuterium fraction measured from \hcope isotopologues. \cite{Redaelli24}, by applying the method on synthetic observations based on {magneto-hydrodynamic} simulations, found that it is usually accurate within a factor of $2-3$. The method was applied in the high-mass regime by \cite{Sabatini20} using observations made with the Atacama Pathfinder EXperiment 12-meter submillimeter telescope (APEX; \citealt{Gusten06}) and the Institut de Radioastronomie Millimétrique (IRAM) 30m single-dish telescope. Their results present \crir values in the range $(0.7-6)\times 10^{-17} \rm \, s^{-1}$, in agreement with theoretical predictions. \cite{Sabatini23} used data from the Atacama Large Millimeter and sub-millimeter Array (ALMA), obtaining \crir maps at high resolution in two high-mass clumps. Their findings are in agreement with the most recent predictions of cosmic-ray propagation and attenuation \citep{padovani22}. \cite{Socci24} used the same method to estimate \crir on parsec scales towards the Orion Molecular Clouds (OMC-2 and OMC-3) by combining IRAM-30m, \textit{Herschel} and Planck observations. All these studies targeted high-mass star-forming regions, and no dedicated analysis based on the \ohhdp methodology in the low-mass regime has been performed so far.}
 \par
 From the theoretical point of view, significant efforts have been made to constrain the \crir in the ISM starting from models of CR propagation and attenuation, taking into account the continuous energy losses that they suffer when interacting with the gas. \cite{Padovani09} computed the local CR spectrum using the continuous slowing-down approximation, and \cite{Padovani18} further expanded the analysis to a higher column density regime. \cite{padovani22} introduced three main models for the relation between \crir and the total gas column density $N \rm (H_2)$, based on three distinct assumptions on the spectral slope $\alpha$ of low-energy protons: model ``low'' $\mathscr{L}$, which uses $\alpha = 0.1$, reproduces the most recent Voyager data \citep{Cummings16, Stone19}; model ``high'' $\mathscr{H}$ adopts $\alpha = -0.8$ and reproduces the high \crir measured in the diffuse medium, and might not be needed anymore in light of the recent results of \cite{Obolentseva24}; the ``up-most'' model ($\mathscr{U}$), with $\alpha = -1.2$, which can be considered an upper limit to the CR ionisation rate by Galactic CRs. 
\par
The use of different methods, based on distinct assumptions and observables, makes it hard to compare the observational results in a statistical way and to interpret them in the framework of current theoretical models. This work aims to produce a uniform sample of measurements of \crir in 20 dense cores, in order to infer the typical \crir value in starless/prestellar sources and to study potential environmental effects, such as proximity to CR sources. We use new high-sensitivity observations obtained with the APEX telescope and adopt the methodology of \cite{Bovino20}. \par
The paper is organised as follows. Section~\ref{sec:sample} describes the selection of the targets. Section~\ref{sec:obs} presents the observational data. The analysis is reported in Sect.~\ref{sec:analysis}. First, we describe how we infer the gas properties in terms of density and temperature from the available \textit{Herschel} maps (Sect.~\ref{sec:H2_tdust}); then we compute the column densities of the targeted molecular species (Sect.~\ref{sec:Ncols}); in Sect.~\ref{sec:unc} we comment on the main sources of uncertainty in our methodology. Section~\ref{sec:fd_rd} reports the estimates of the CO depletion factor and deuteration fraction, while Sect.~\ref{sec:CRIR} presents the results concerning \crir. In Sect.~\ref{sec:xe} we assess the cores' ionisation degree and dynamical states. We discuss the results in Sect.~\ref{sec:disc}. A summary is presented in Sect.~\ref{sec:summary}.
\begin{table*}
\centering
    \renewcommand{\arraystretch}{1.2}

\caption{Properties of the sample targeted by this work. The gas column density and dust temperature are taken from the maps generated from the \textit{Herschel} continuum data. The last column indicates the corresponding object in the HGBS catalogues (when present), and the distance offset to our pointing. \label{tab:cores}}
\begin{tabular}{cccccccc}
\hline
\hline
Core    &   \multicolumn{2}{c}{Coordinates}  &   Distance\tablefootmark{a}&  $N\rm(H_2)$\tablefootmark{b}& \tdust\tablefootmark{b}  &   $L$\tablefootmark{c}    &  HGBS catalogue object \\
        &   R.A. (J2000)      & Dec.(J2000)     & pc    &$10^{22}\,\rm cm^{-2}$&K&pc        &       \\
\hline
\multicolumn{8}{c}{Taurus} \\
Tau 410	&	$\rm 04^{h}18^{m}32.99^{s}$	&	$\rm +28^{\circ}28^{m}29.0^{s}$	&	130	&	1.8	&	11.4	&	0.096-0.149&Same		 \\ 
Tau 420	&	$\rm 04^{h}18^{m}40.32^{s}$	&	$\rm +28^{\circ}23^{m}16.0^{s}$	&	130	&	4.0	&	12.8	&	0.064-0.120&Same		 \\ 
TMC1-C\tablefootmark{d}	&	$\rm 04^{h}41^{m}34.31^{s}$	&	$\rm +26^{\circ}00^{m}40.0^{s}$	&	140	&	1.9	&	11.3	&	0.129-0.157&-		 \\ 
L1544	&	$\rm 05^{h}04^{m}17.21^{s}$	&	$\rm +25^{\circ}10^{m}42.8^{s}$	&	170	&	11.6	&10.2	&	0.073-0.125&	-	 \\ 
\multicolumn{8}{c}{Isolated cores} \\
L183	&	$\rm 15^{h}54^{m}08.56^{s}$	&	$\rm -02^{\circ}52^{m}49.0^{s}$	&	110	&	5.9	&	9.6		&	0.052-0.077&	-	 \\ 
L429	&	$\rm 18^{h}17^{m}05.53^{s}$	&	$\rm -08^{\circ}13^{m}29.9^{s}$	&	200	&	5.1	&	9.9		&	0.065-0.109&	-	 \\ 
L694-2	&	$\rm 19^{h}41^{m}05.03^{s}$	&	$\rm +10^{\circ}57^{m}02.0^{s}$	&	250	&	4.3	&	9.8		&	0.077-0.111&	-	 \\ 
\multicolumn{8}{c}{Ophiuchus} \\
Oph 1	&	$\rm 16^{h}31^{m}57.63^{s}$	&	$\rm -24^{\circ}57^{m}35.7^{s}$	&	140	&	3.8	&	13.4	&	0.067-0.141&Oph 410+412\tablefootmark{e} \\
Oph 2	&	$\rm 16^{h}31^{m}39.94^{s}$	&	$\rm -24^{\circ}49^{m}50.3^{s}$	&	140	&	5.2	&	13.3	&	0.075-0.115&Oph 387 ($24''$) \\ 
Oph 3	&	$\rm 16^{h}27^{m}33.24^{s}$	&	$\rm -24^{\circ}26^{m}24.1^{s}$	&	140	&	7.8	&	12.5	&	0.085-0.110&Oph 246 ($7.8''$)		 \\ 
Oph 4	&	$\rm 16^{h}27^{m}12.77^{s}$	&	$\rm -24^{\circ}29^{m}40.3^{s}$	&	140	&	5.6	&	13.9	&	0.084-0.142&Oph 196 ($9.2''$)		 \\ 
Oph 5	&	$\rm 16^{h}27^{m}15.28^{s}$	&	$\rm -24^{\circ}30^{m}30.1^{s}$	&	140	&	5.4	&	13.9	&	0.081-0.125&Oph 201 ($11''$)		 \\ 
Oph 6	&	$\rm 16^{h}27^{m}19.99^{s}$	&	$\rm -24^{\circ}27^{m}17.6^{s}$	&	140	&	4.7	&	13.8	&	0.117-0.155&Oph 215 ($5.7''$)		 \\ 
Oph D	&	$\rm 16^{h}28^{m}28.56^{s}$	&	$\rm -24^{\circ}19^{m}25.0^{s}$	&	140	&	1.7	&	13.3	&	0.076-0.108&Oph 316 ($11''$)		 \\ 
\multicolumn{8}{c}{Corona Australis} \\
CrA 038	&	$\rm 19^{h}01^{m}46.10^{s}$	&	$\rm -36^{\circ}55^{m}35.7^{s}$	&	150	&	6.2	&	13.7	&	0.078-0.109&Same		 \\ 
CrA 040	&	$\rm 19^{h}01^{m}47.28^{s}$	&	$\rm -36^{\circ}56^{m}39.8^{s}$	&	150	&	4.6	&	15.5	&	0.115-0.144&Same		 \\ 
CrA 044	&	$\rm 19^{h}01^{m}54.45^{s}$	&	$\rm -36^{\circ}57^{m}48.9^{s}$	&	150	&	10.5	&18.2	&	0.044-0.074&Same		 \\ 
CrA 047	&	$\rm 19^{h}01^{m}55.86^{s}$	&	$\rm -36^{\circ}57^{m}46.9^{s}$	&	150	&	9.9	&	19.0	&	0.044-0.071&Same		 \\ 
CrA 050\tablefootmark{f}	&	$\rm 19^{h}01^{m}58.94^{s}$	&	$\rm -36^{\circ}57^{m}09.9^{s}$	&	150	&	3.5	&	18.0	&	0.107-0.132&Same		 \\ 
CrA 151	&	$\rm 19^{h}10^{m}20.17^{s}$	&	$\rm -37^{\circ}08^{m}27.0^{s}$	&	150	&	4.1	&	11.2	&	0.041-0.076&Same		 \\ 
\hline
\end{tabular}
\tablefoot{
\tablefoottext{a}{Distance references: Ophiuchus, \cite{OrtizLeon18}; Taurus, \cite{Galli19}; Corona Australis, \cite{Galli20}; L183, L429, L694-2, \citet[][and references therein]{Redaelli18}.} \\
\tablefoottext{b}{Values towards the {cores' coordinates}, averaged in a $37''$ beam.} \\
\tablefoottext{c}{Path length along the line-of-sight from where the molecular emission arises in Eq.~\ref{eq:crir_observables}. The two values correspond to the equivalent diameter of the region encompassing the 60\% and 40\% contours of the $N \rm (H_2)$ peak value, as described in Sect.~\ref{sec:CRIR}.}\\
\tablefoottext{d}{The TMC1-C pointing is offset by $\approx 60''$ compared to the one of \cite{Caselli08}.} \\
\tablefoottext{e}{Oph 1 is found roughly halfway between the centres of Oph 410 ($22''$ to the North-West) and Oph 412 ($18''$ to the South-West) from the HGBS catalogue. For the \dcop (4-3) spectrum towards this object, we have taken the average of the two spectra. Their intensities differ by $\sim 5$\%.}\\
\tablefoottext{f}{CrA 050 contains a detected YSO within its boundaries (source SMM2, \citealt{Sicilia-Aguilar13}). This is a young, Class I object.}
}

\end{table*}
\section{The source sample\label{sec:sample}}
The sample analysed in this work is constituted of 20 starless and/or prestellar cores embedded in nearby (distance $\lesssim 250 \, $pc) low-mass star-forming regions. Table~\ref{tab:cores} lists their name, coordinates, and distances. To take advantage of observations already available, we included the six Ophiuchus cores from \cite{Bovino21}, who reported the detection of \olineh with APEX. We added the six objects belonging to the sample of \cite{Caselli08} where \ohhdp was detected using the Caltech Submillimeter Observatory (CSO): L1544 and TMC1-C in Taurus, Ophiuchus D in Ophiuchus, L183, L694-2, and L429. \par
To complete the target list, we extracted 8 sources from the sample of cores identified by the \textit{Herschel} Gould Belt Survey (HGBS\footnote{All data available at \url{http://gouldbelt-herschel.cea.fr/archives}.}; \citealt{Andre10}. See \citealt{Kirk13, Marsh16, Bresnahan18} for the core catalogues). Caselli et al. (in prep.) selected a sample of 40 objects from these catalogues, chosen to have high central volume and column densities ($n(\rm H_2) > 10^5 \, cm^{-3}$, $N(\rm H_2) > 5 \times 10^{22} \, cm^{-2}$) and presented follow-up observations of high-$J$ transitions of $\rm N_2H^+$ and $\rm N_2D^+$, obtained with APEX. Further publications focus on two specific cores {(CrA 151, \citealt{Redaelli25}; Oph 464, also known as IRAS16293E, \citealt{Spezzano25})}. We drew eight from this catalogue of starless sources in Corona Australis and Taurus.
The seven cores in Ophiuchus also have correspondence to HGBS objects from \cite{Ladjelate20}, even though the coordinates of the centres are offset by $5-20''$ when compared to our pointings  (see Table~\ref{tab:cores} for more details). \par
After the observational campaign, we noted that the target CrA 050 contains a detected Class I Young Stellar Object (YSO) within the core boundaries \citep{Sicilia-Aguilar13}. We decided to keep the core in the sample, however, to investigate the ionisation rate in a young protostellar envelope. {Furthermore, \cite{Redaelli25} discussed that CrA 151 might also contain a very young protostar, but the evidence is so far inconclusive.}

\section{Observations \label{sec:obs}}
\subsection{APEX data}
The molecular line observations were performed with the APEX single-dish antenna as part of two projects, IDs M9505A\_111 and M9503A\_112 (PI: E. Redaelli). In March and June 2023, we observed the sources in Ophiuchus, Corona, and the isolated ones (L429, L694-2, L183). The Taurus cores were instead observed in the second half of 2023. \par
We used four spectral setups in total. Setup 1 covered the \olineh using the LAsMA receiver \citep{Gusten08}, and it was not re-observed in the six Ophiuchus cores of \cite{Bovino21}. Setup 2 covered simultaneously the \dcop (3-2) and \cdo (2-1) lines using the receiver nFLASH230. Setup 3 was dedicated to the $\rm HC^{18}O^+$ and \htcop (2-1) lines using the receiver SEPIA180 \citep{Belitsky18a, Belitsky18b}. Setup 4 covered simultaneously the \hcope and \htcop (3-2) lines using nFLASH230. 
 \par
\begin{figure*}
    \centering
    \includegraphics[width=\linewidth]{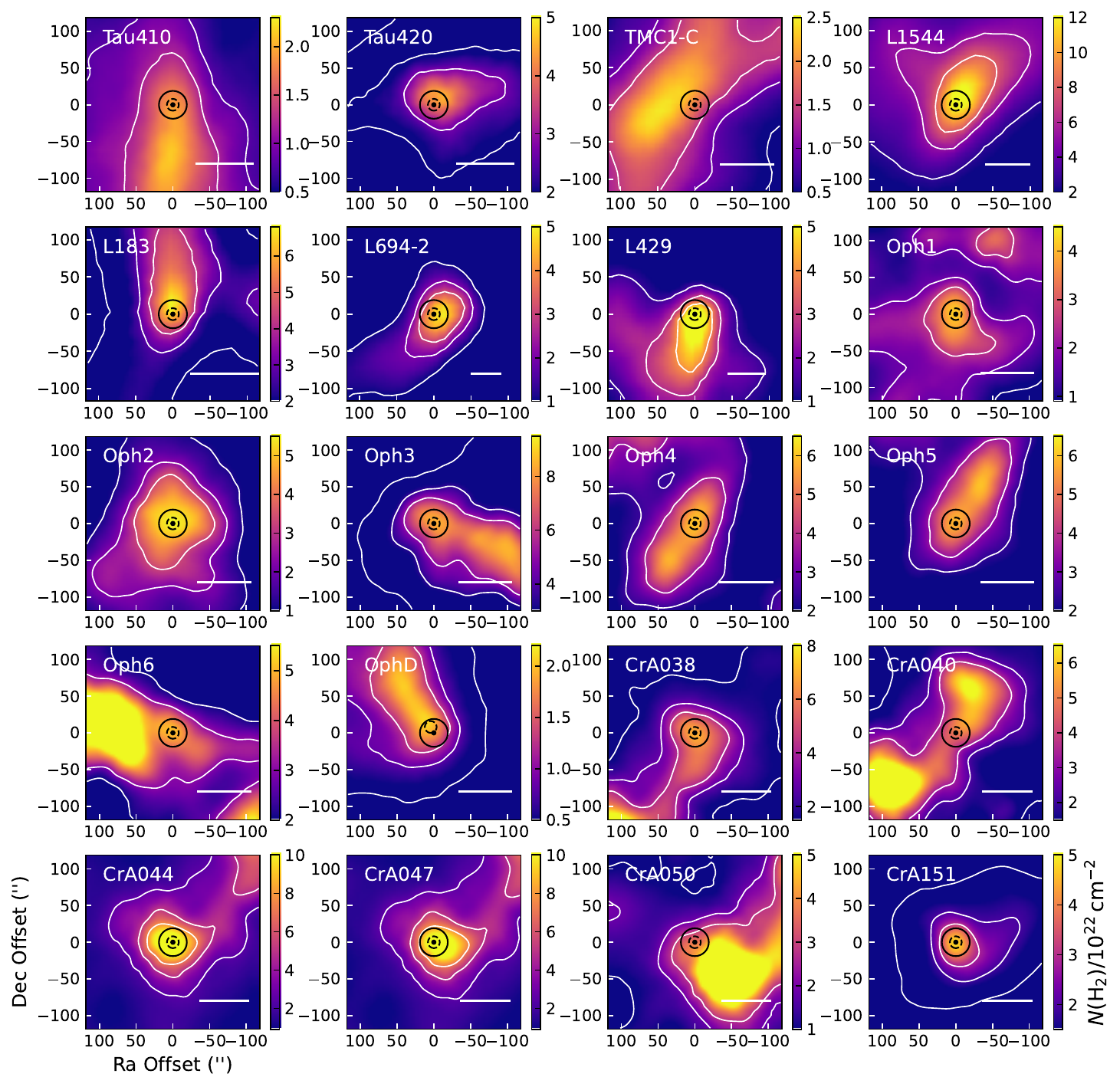}
    \caption{$N\rm (H_2)$ maps towards each core in the sample (labelled in the top-left corner of every panel). The scalebar shown in the bottom-right corner represents a length of $0.05\,\rm pc$. The white contours show the 20, 40, and 60\% levels of the peak value within the central \textit{Herschel} beam. The solid circles show the {APEX} pointing and the $N\rm (H_2)$ map beam size, whilst the dashed circles show the beam sizes and pointings of APEX for the \ohhdp line. Note the small shift present between the two positions for Oph D, where the \ohhdp beam is not the central LAsMA beam, but one of the external ones (see Main Text). The shift is however smaller than the continuum and APEX resolutions.}
    \label{fig:continuum}
\end{figure*}

 For all the setups, we used the APEX FFTS spectrometer at the highest resolution of 64$\,$kHz, which translates into a velocity resolution of $0.05-0.11\,$\kms. The observations were performed as single-pointings towards the {cores' coordinates listed in Table~\ref{tab:cores} (see also Fig.~\ref{fig:continuum})}. In the case of LAsMA, a multi-beam receiver, we set the central beam in the core centre. The only exception is Ophiuchus D, where instead the core centre was located in one of the six external beams, to ensure the same spatial coverage as previous observations. We used the position switching observation mode, with the off-position hand-picked usually 0.5$^\circ$ away from the on position. 
 \par
 In addition, we analysed the published \olineh data towards the cores Oph 1 - Oph 6. We refer to \cite{Bovino21} for an exhaustive description of the observations. We also used \dcop (4-3) observations obtained with APEX, which are available for all the targets except L429, L183, L694-2, TMC1-C, and L1544 (from Caselli et al., in prep.). \par
 We reduced the data using the \textsc{class} package from the \textsc{gildas} software\footnote{Available at \url{http://www.iram.fr/IRAMFR/GILDAS/}.} \citep{Pety05}. In particular, we subtracted baselines (usually a first-order polynomial) in every single scan, and then averaged them. The intensity scale was converted in main beam temperature $T_\mathrm{MB}$ assuming the forward efficiency $F_\mathrm{eff} = 0.95$ and computing the main-beam efficiency $\eta_\mathrm{MB}$ at each frequency based on the data tabulated at \url{https://www.apex-telescope.org/telescope/efficiency/index.php}. Relevant information on the targeted transitions and their observational parameters is given in Table~\ref{tab:lines}, including the angular resolution ($\theta_{\rm MB}$) of the observations, the main-beam efficiency values, and the velocity resolution $\Delta V_\mathrm{ch}$.
\begin{table*}[!h]
\centering
    \renewcommand{\arraystretch}{1.2}
\caption{Properties of the targeted lines. We list the frequency ($\nu$), angular resolution, main-beam efficiency, and velocity resolution. In the last three columns, we report the spectroscopic constants used in the column density computation (see Sect.~\ref{sec:Ncols}). They are taken from the CDMS catalogue \citep{Muller05}.}
\label{tab:lines}
\begin{tabular}{cccccccc}
       \hline \hline
Transition & $\nu$& $\theta_\mathrm{MB}$ &  $\eta_\mathrm{MB}$&$\Delta V_\mathrm{ch}$ & $A_\mathrm{ul}$ & $g_\mathrm{u}$ & $E_\mathrm{u}$  \\
 		& MHz&		$''$			&			& \kms	& $\times 10^{-5} \rm s^{-1}$	&		& K			 \\		
		\hline	
 $\rm HC^{18}O^+$ (2-1)       & 170322.626       & 35.4 &	0.86&0.11	               & 34.99 & 5          & 12.26  	          \\
 
 $\rm H^{13}CO^+$ (2-1)                     & 173506.700             &34.7 &	0.86&0.11& 36.99 & 5 		&	12.49	            \\
 
 $\rm DCO^+$ (3-2)           & 216112.582       &27.9	&	0.82&  0.08           & 76.58 & 7		&	12.49	                  \\
  
$\rm   C^{18}O$ (2-1)        & 219560.350          &27.5 & 0.82 & 0.08               & 0.0601  &5 & 15.80                              \\

$\rm H^{13}CO^+$ (3-2)      & 260255.339          & 23.2& 0.78& 0.07                & 133.75  &7 & 24.98                             \\

$ \rm HCO^+ $ (3-2)         & 267557.626          & 22.5& 0.78 & 0.07                 & 145.30  &7 & 25.68                       \\

  $\rm DCO^+$ (4-3)             & 288143.858             & 20.9 & 0.76 & 0.06           & 188.23  &9 & 34.57                           \\

 $\rm oH_2D^+ (1_{1,0}-1_{1,1}) $   & 372421.356   & 16.2& 0.69& 0.05\tablefootmark{a}        & 10.96  &9 & 17.90                            \\
 \hline
\end{tabular}
\tablefoot{
\tablefoottext{a} {The velocity resolution of the archival APEX data for \olineh in cores Oph 1-6 is 0.2$\,$\kms \citep{Bovino21}.}
}
\end{table*}
\subsection{Continuum data\label{sec:data_cont}}
The methodology we applied (see Sect.~\ref{sec:analysis}) requires knowledge of the gas total column density and temperature. We have used available maps of $N\rm (H_2)$ and dust temperature \tdust derived from the spectral-energy-distribution (SED) fit of the dust continuum emission detected with \textit{Herschel}. In particular, for the cores in Ophiuchus, Corona Australis, and Taurus (except for L1544) we have used the maps provided by the HGBS survey\footnote{The detailed references are: \cite{Bresnahan18} for Corona; \cite{roy14} and \cite{Ladjelate20} for Ophiuchus; \cite{Palmeirim13} for Taurus.}. For the remaining cores, we have used the maps published in \citet[for L1544]{Spezzano17} and \citet[for L429, L694-2, and L183]{Redaelli18}. The HGBS survey provides H$_2$ column density map at the high angular resolution of $18.2''$ (cf. \citealt{Palmeirim13}). In this work, however, we used the products at the larger angular resolution of the SPIRE/500$\,\mu$m wavelength ($\approx 37''$). This is closer to the APEX resolution at $\sim 2\,$mm, and it is the one available for L183, L429, L694-2, and L1544. 

\section{Analysis and Results\label{sec:analysis}}

To compute \crir, we applied the analytical formula of \cite{Bovino20}, which was derived from first principles and has the advantage of being model-independent. {We refer to that paper and to \cite{Redaelli24} for a complete description of the method assumptions and limitations. The approach} is based on the detection of the first deuterated form of $\rm H_3^+$, i.e. \ohhdp:
 \begin{equation}
 \zeta_2 = k_\mathrm{CO}^\mathrm{oH_3^+} \frac{1}{3}\times X(\mathrm{CO}) \times \frac{N(\mathrm{oH_2D^+})}{R_\mathrm{D}}\frac{1}{L}    \; ,
 \label{eq:crir_observables}
 \end{equation}
where the deuterium fraction of \hcope is $R_\mathrm{D} = N(\mathrm{DCO^+} )/ N(\mathrm{HCO^+})$, the CO abundance (with respect to H$_2$) is $X(\mathrm{CO})  =  N(\mathrm{CO})/N(\mathrm{H_2})$, and $N(\mathrm{oH_2D^+})$ is the column density of \ohhdp. $k_\mathrm{CO}^\mathrm{oH_3^+}$ is the destruction rate of oH$_3^+$ by CO, assumed here to be the main destruction path for H$_3^+$, while $L$ is the path length over which the column densities are estimated. The $k_\mathrm{CO}^\mathrm{oH_3^+}$ rate is temperature dependent, and we have computed it in each core assuming $T_\mathrm{gas} = T_\mathrm{dust}$, hence assuming dust and gas coupling (expected to be efficient at $n \gtrsim 10^{4-5}\, \rm cm^{-3}$, \citealt{Goldsmith01}). \cite{Sokolov17} showed that the difference in the two temperatures can be of the order of $2\,$K, supporting that this is a good approximation. The rate value is taken from the KIDA\footnote{ \url{https://kida.astrochem-tools.org}.} database \citep[][]{Wakelam12}, and it presents a modest decrease in the temperature range $8-20\, \rm K$. \par
We used the optically thinner \cdo to infer the CO abundance, assuming $X(\mathrm{CO}) = 557 \times X (\rm C^{18}O)$ \citep{Wilson99}, avoiding opacity issues. For the same reason, the \hcope deuteration level has been computed from \dcop and the rarer \htcop isotopologue, assuming the isotopic ratio of $\rm ^{12}C/^{13}C = 68$ \citep{Wilson99}. In fact, the \hcope (3-2) transition presents in most sources double-peak profiles with asymmetries, indicative of self-absorption, and it is not suitable to infer the molecular abundance. The $\rm HC^{18}O^+$ (2-1) is detected only in 60\% of the sample, and we do not include it in the analysis. However, we use it later to check the carbon isotopic ratio in the sample (see Sect.~\ref{sec:unc}).

\subsection{Total gas column density and dust temperature \label{sec:H2_tdust}}
To assign to each core a value of total gas column density $N\rm (H_2) $ and dust temperature $T_\mathrm{dust}$, we have taken the maps listed in Sect.~\ref{sec:data_cont}. We cut square regions of $4' \times 4'$ around the cores' central coordinates. The maps are shown in Fig.~\ref{fig:continuum}. We measured the average \tdust and $N\rm (H_2)$ in one \textit{Herschel} beam ($37''$) around the centre. The values are listed in Table~\ref{tab:cores}. \par
The cores span the temperature range of $9-20 \, \rm$K, with a mean value of $13\, \rm K$. The warmest sources all belong to the Corona Australis region (CrA 038, 040, 044, 047, and 050), in particular in the area of the Coronet cluster, a well-known active star-forming region \citep[cf.][]{Chini03,  Sicilia-Aguilar13, Sabatini24}. The column density values range from a minimum of $1.7 \times 10^{22} \, \rm cm^{-2}$ (Oph D), to
a maximum of $1.16 \times 10^{23} \, \rm cm^{-2}$ (L1544), i.e. approximately one order of magnitude of variation. 
\subsection{Molecular column densities\label{sec:Ncols}}
The estimate of \crir depends on the measurements of column densities of four distinct tracers in Eq.~\eqref{eq:crir_observables}. We adopted the constant excitation temperature approach (C-TEX, \citealt{Caselli02b}) {to} fit the observed spectra and derive {the column density (\ncol) for each species}. The {fitting} procedure is implemented using the \textsc{python} package \textsc{pyspeckit} \citep{Ginsburg11, Ginsburg22}, {with} the spectroscopic constants listed in Table~\ref{tab:lines}. The partition function of \dcop is taken from \cite{Redaelli19a}, of \cdo, \hcope, \htcop, and $\rm HC^{18}O^+$ from the CDMS database, and of \ohhdp from our own calculation based on the energy data provided by CDMS\footnote{ The partition function of \ohhdp at the temperature $T$ is computed using $Q = \sum_i g_{u,i}\exp \left( -\frac{E_{u,i}}{k_B T} \right)$, where $g_{u,i}$ and $E_{u,i}$ are the degeneracy and energy of the upper $i$ level.}. We highlight that the transitions from D-bearing species (\dcop, \ohhdp) present hyperfine splitting due to the nuclear spin of the deuterium atom. However, the hyperfine structure is usually unresolved given the spectral resolution of our observations \citep[cf.][]{Caselli05}. Since we are not interested in high-precision kinematic measurements in this work (where including the hyperfine structure would be relevant), we neglect it. \par
The code fits the observed spectra using four free parameters: the line centroid velocity $V_\mathrm{lsr}$, the {line velocity dispersion} $\sigma_\mathrm{V}$, the molecular column density \ncol and the excitation temperature \tex. When two transitions from the same species are available, the method assumes that \tex is equal for all transitions. This allows fitting \tex and \ncol simultaneously. We estimated the column density of \htcop with this approach in the whole sample, except for L1544, Tau 410, and TMC1-C, where \htcop (3-2) is not detected\footnote{In L1544, a tentative detection of \htcop (3-2) was found, but the low S/N ($\sim 3$) prevents the fit from converging.}. Similarly, the (3-2) and (4-3) transitions of \dcop are available for 14 out of 20 targets. \par
\begin{figure}[!t]
    \centering
    \includegraphics[width=\linewidth]{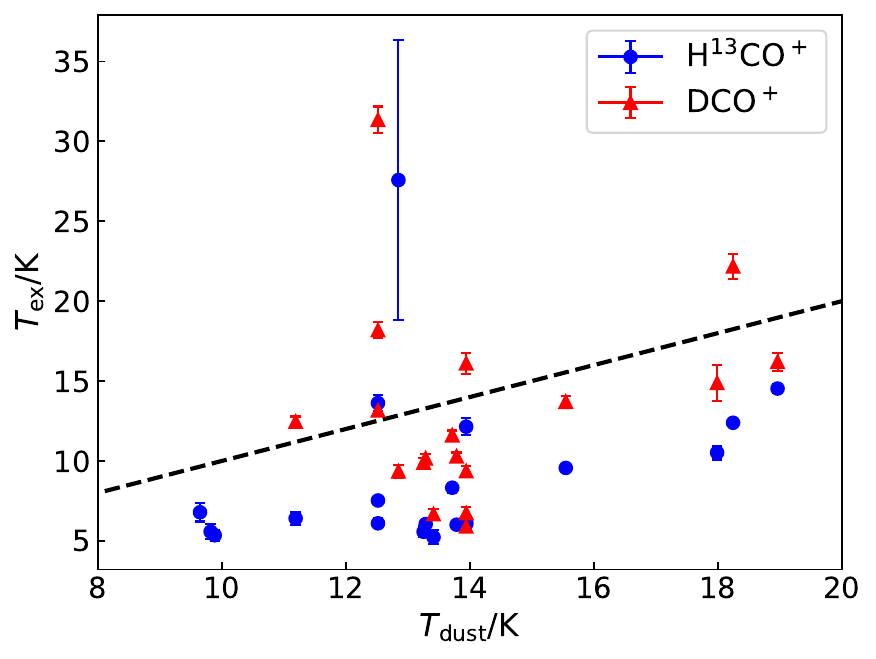}
    \caption{\tex of \htcop (blue circles) and \dcop (red triangles) for all cores where two transitions of the species are available, as a function of \tdust values. The dashed black line shows the 1:1 relation. For cores where two velocity components are identified and fitted, we show the parameters of both.  \label{fig:Tdust_Tex}}
\end{figure}

In Fig.~\ref{fig:Tdust_Tex} we compare the \tex values obtained for \dcop and \htcop in all the cores where this has been possible, also with the \tdust values obtained as described in Sect.~\ref{sec:H2_tdust}. In general, the excitation temperatures are lower than the corresponding \tdust value by 6\% for \dcop and by 34\% for \htcop. This indicates that the lines are generally sub-thermally excited, i.e. the collisional excitations are not efficient enough to counterbalance the radiative de-excitations, or that the \tdust is overestimating the gas temperature, because of line-of-sight averaging (see \citealt{Sokolov17}). The first effect is expected as these lines have high critical densities, which can be higher than the average {densities} of the cores. Interestingly, the \dcop \tex is on average higher than the \htcop one: considering all cases where we have estimates for both, $ T \mathrm{_{ex} (DCO^+)} = 1.4 \times  T \mathrm{_{ex} (H^{13}CO^+)} $. We struggle to find an explanation for this behaviour. The fact that the former is estimated using higher-$J$ transitions than the latter (hence with higher critical densities) should point in the opposite direction. Furthermore, \htcop is not systematically less abundant than \dcop, which could justify a difference in excitation conditions. 
The average optical depth of these lines is consistently $\tau < 1$, which rules out opacity issues (see Table~\ref{tab:fit_params}). { Figure~\ref{fig:hcop_dcop} shows the comparison of the centroid velocity and velocity dispersion values in the whole sample. The $V_\mathrm{lsr}$ values are remarkably consistent within each other. The $\sigma_\mathrm{V}$ values present a higher scatter, and in 68\% of sources the \dcop lines are narrower than the \htcop ones.} A possible explanation is that the \htcop lines are tracing slightly lower densities than the \dcop lines, as \textit{i)} deuteration is expected to increase towards the centre and \textit{ii)} the angular resolution of \htcop data is worse than the \dcop one, which could cause more lower density material to be intercepted by the \htcop emission.
\begin{figure*}[!h]
    \centering
    \includegraphics[width=0.8 \textwidth]{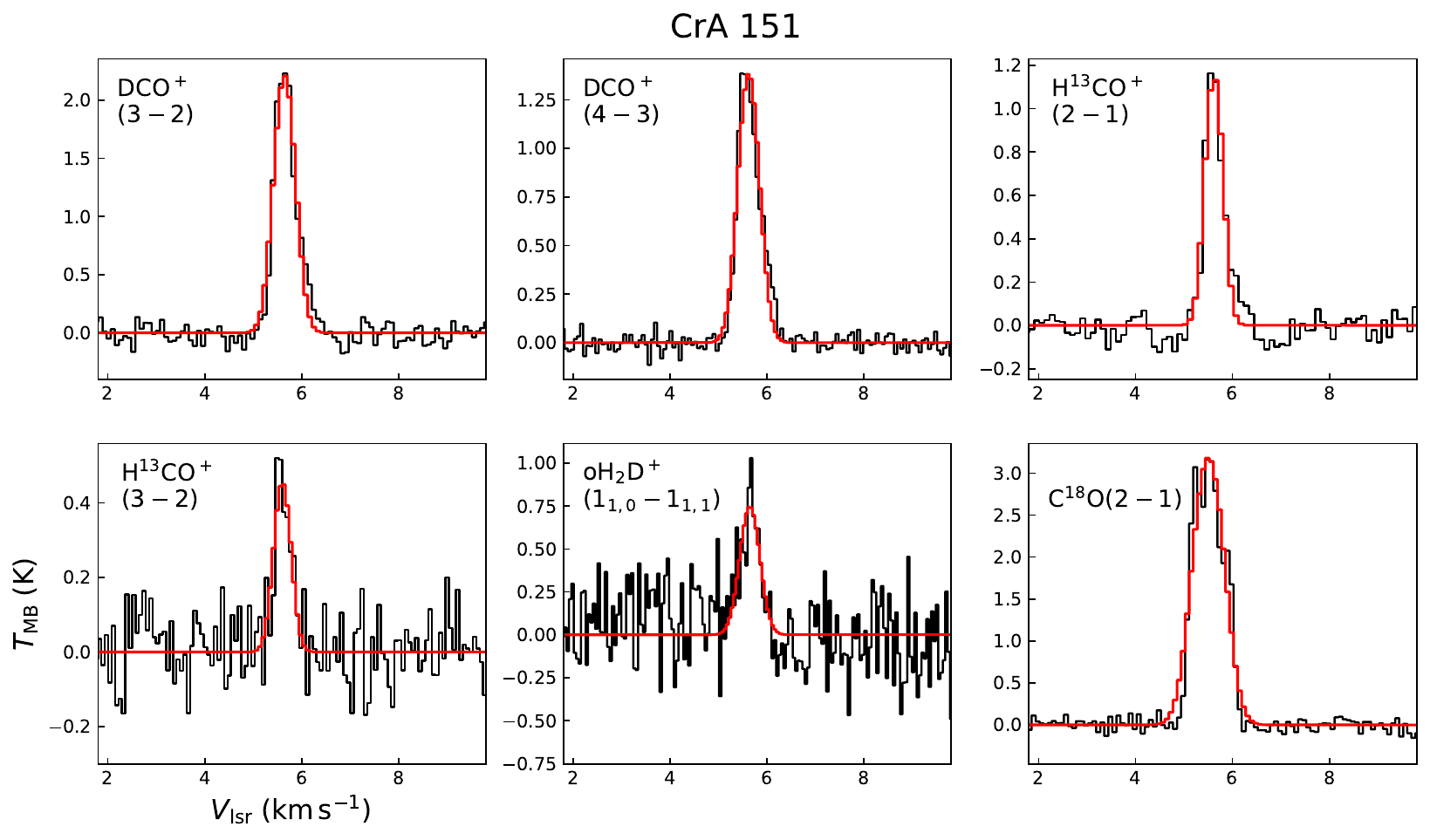}
    \caption{The black histograms show the spectra collected towards CrA 151. The transitions are labelled in the top-left corner of each panel. The red histograms show the best-fit solution of the spectral fit performed as described in Sect.~\ref{sec:Ncols}.\label{fig:cra_151}}
\end{figure*}
\par
For the species with one detected line, \tex and \ncol are degenerate parameters, and the former must be assumed to estimate the latter. We highlight that this choice is arbitrary, as there is no general prescription to estimate \tex. We proceed as follows. For \htcop and \dcop, in those cases where only one line is detected or available, we compute \tex based on the \tdust values, using the corrective factors computed above (0.94 for \dcop and 0.66 for \htcop). For \ohhdp we adopt the same \tex as of \dcop. These values are in the range $6-22\, $K and they compare well with previous studies of the same molecule \citep{Vastel06, Caselli08, Redaelli21, Redaelli22}. For \cdo we assume $\text{\tdust} = \text{\tex}$, because this is an abundant species with low critical density and it is likely to be thermalised. \par
\begin{figure*}[!h]
    \centering
    \includegraphics[width=0.8 \textwidth]{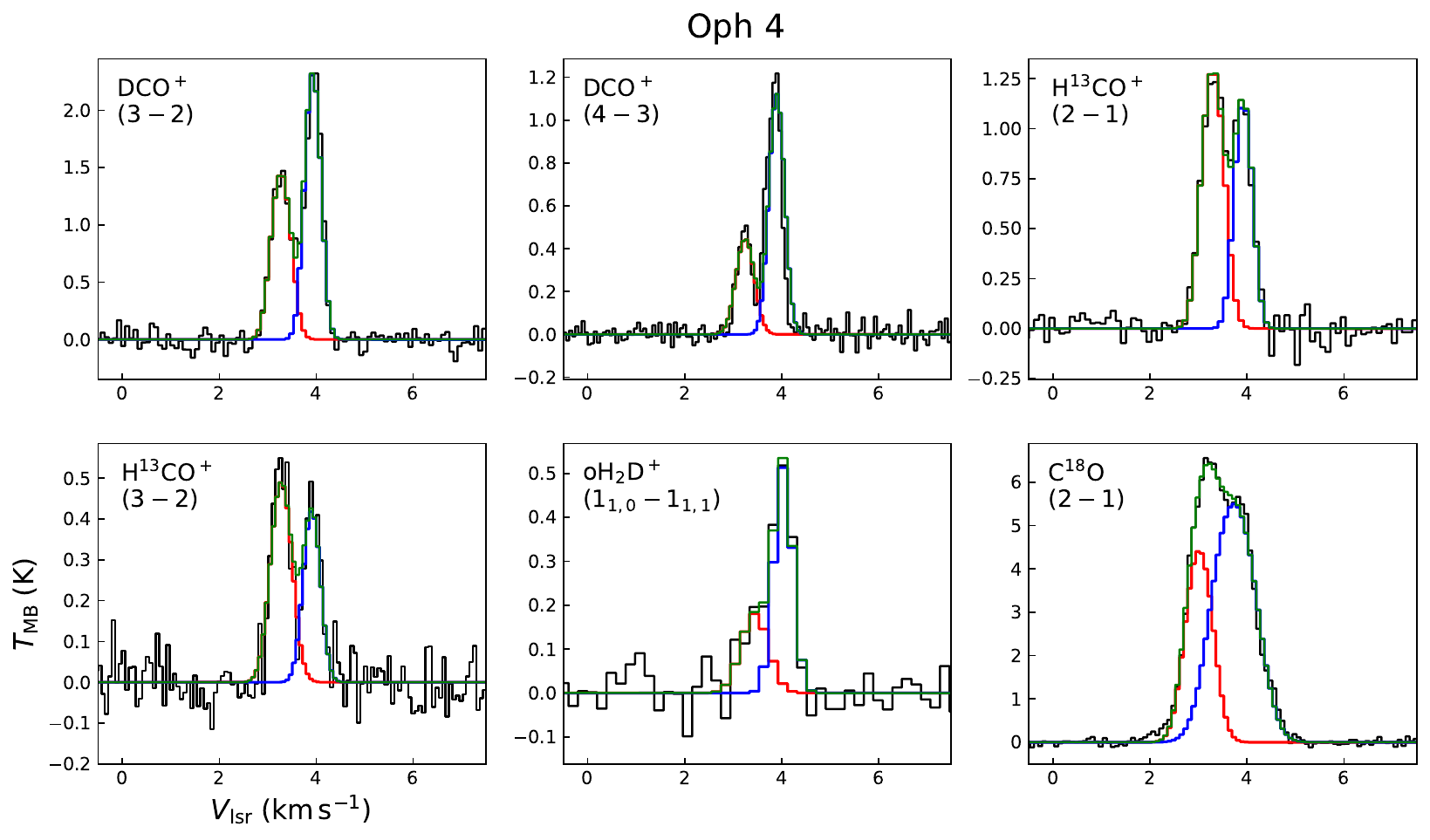}
    \caption{Same as Fig.~\ref{fig:cra_151}, but for Oph 4. In this core, two velocity components are seen in all the transitions, and we fit them separately. The red/blue curves show the individual best-fit solutions, whilst the green curve represents the total fit.\label{fig:oph4}}
\end{figure*}

Figure~\ref{fig:cra_151} showcases an example of the best-fit solutions found towards CrA 151. The lines are well reproduced by the spectral modelling. For a few sources in the sample, however, the line profiles deviate from single Gaussian components. This is the case, for instance, for Oph 4, shown in Fig.~\ref{fig:oph4}, where two clear velocity components (one at $\sim 3.4\,$\kms and one at $\sim 3.9\,$\kms) are visible. For the cases where two velocity components can be seen in all transitions, we perform the fit routine previously described with two components, deriving two sets of molecular column densities for the same target. We then sum the two contributions to obtain one column density value for each species in each core. This is necessary, as the \tdust and total column density values obtained from the continuum maps cannot distinguish among distinct velocity components. This approach is used for Oph 3 and Oph 4. For other cores where the multiple components are not visible in all tracers or they are too overlapped for the fit to converge, we adopt the single-component procedure, in line with our goal to infer an average \crir value across each core. Table~\ref{tab:col_dens} summarises the \tex and \ncol values obtained as just described. The complete set of figures for all cores is available in Appendix \ref{app:full_figures}. Table \ref{tab:fit_params} lists the derived best-fit values for the remaining free parameters, {line velocity dispersion} and centroid velocity.

\begin{table*}[!h]
\centering
    \renewcommand{\arraystretch}{1.2}

\caption{Molecular excitation temperatures and column densities obtained {by} fitting the observed spectral lines. When the \tex value is presented without uncertainties, its value has been assumed and fixed in the fit (see the Main Text for more details). }
\label{tab:col_dens}
\begin{tabular}{c|cc|cc|cc|cc}
\hline \hline
Source  &   \multicolumn{2}{c|}{\htcop} & \multicolumn{2}{c|}{\dcop} & \multicolumn{2}{c|}{\ohhdp} &\multicolumn{2}{c}{\cdo}\\
        &\tex&  $N_\mathrm{col}$&\tex&  $N_\mathrm{col}$&\tex&  $N_\mathrm{col}$&\tex&  $N_\mathrm{col}$ \\
        
      & K&  $\rm 10^{11}\,cm^{-2}$& K&  $10^{11}\, \rm cm^{-2}$& K&  $\rm 10^{13}\,cm^{-2}$& K&  $\rm 10^{15}\, cm^{-2}$ \\
\hline 
Tau 410	&	$7.5$	    &$1.28\pm0.10$	& $10.8$        &	$0.88\pm0.12$	&	$10.8$	&	$0.18\pm0.03$	&	$11.4$	&	$0.816\pm0.019$	\\
Tau 420	&	$28\pm 9 $&$0.89\pm0.08$	& $9.4\pm0.4$  &	$6.0\pm0.5$	    &	$9.4$	&	$0.24\pm0.04$	&	$12.8$	&	$1.45\pm0.02$	\\
TMC1-C	&	$7.5$	    &$1.18\pm0.10$	& $10.7$	     &	$1.43\pm0.17$	&	$10.7$	&	$0.070\pm0.013$	&	$11.3$	&	$1.22\pm0.02$	\\ 
L1544	&	$6.7$	    &$2.5\pm0.2$	& $9.6$	         &	$4.87\pm0.18$	&	$9.6$	&	$0.98\pm0.08$	&	$10.2$	&	$1.16\pm0.02$	\\
L183	&	$6.8\pm0.6$	&$1.5\pm0.2$	& $9.1$ 	     &	$5.22\pm0.18$	&	$9.1$	&	$1.28\pm0.08$	&	$9.6$	&	$1.68\pm0.02$	\\
L429	&	$5.4\pm0.3$	&$4.3\pm0.7$	& $9.3 $	     &	$8.9\pm0.3$ 	&	$9.3$	&	$1.35\pm0.06$	&	$9.9$	&	$1.11\pm0.03$	\\
L694-2	&	$5.6\pm0.5$	&$3.3\pm0.7$	& $9.3 $	     &	$5.7\pm0.2$	    &	$9.3$	&	$0.72\pm0.06$	&	$9.8$	&	$1.31\pm0.03$	\\
Oph 1	&	$5.2\pm0.4$	&$3.0\pm0.7$	& $6.7\pm0.3$	 &	$13\pm2$	    &	$6.7$	&	$1.76\pm0.14$	&	$13.4$	&	$2.20\pm0.05$	\\
Oph 2	&	$6.0\pm0.3$	&$7.6\pm0.8$	& $10.2\pm0.3$ &	$17.1\pm0.8$	&	$10.2$	&	$0.46\pm0.03$	&	$13.3$	&	$3.20\pm0.03$	\\
Oph 3\tablefoottext{a}	&	- &$22.5\pm 1.8$	& - &	$26.8\pm1.5$        &	-	&	$0.20\pm0.08$	&	-	&	$6.20\pm0.14$	\\
Oph 4\tablefoottext{a}	&	- &$15.6\pm1.3$	& - &	$28 \pm 3$	    &	-	&	$1.0\pm0.3$	&	-	&	$8.5\pm0.4$	\\
Oph 5	&	$6.3\pm0.2$	&$11.0\pm1.0$	& $5.90\pm0.05$	 &	$119\pm10$	    &	$5.9$	&	$2.34\pm0.18$	&	$13.9$	&	$9.11\pm0.08$	\\
Oph 6	&	$6.0\pm0.2$	&$18.1\pm1.6$	& $10.3\pm0.2$ &	$27.7\pm1.3$	&	$10.3$	&	$0.50\pm0.04$	&	$13.8$	&	$5.81\pm0.04$	\\
Oph D	&	$5.6\pm0.3$	&$8.1\pm1.6$	& $9.9\pm0.3$	 &	$12.8 \pm0.8$	&	$9.9$	&	$0.69\pm0.05$	&	$13.3$	&	$1.67\pm0.03$	\\
CrA 038	&	$8.3\pm0.3$	&$8.0\pm0.5$	& $11.6\pm0.3$ &	$11.1\pm0.5$	&	$11.6$	&	$0.65\pm0.03$	&	$13.7$	&	$7.10\pm0.12$	\\
CrA 040	&	$9.6\pm0.2$	&$21.2\pm0.7$	& $13.7\pm0.4$ &	$14.5\pm0.5$	&	$13.7$	&	$0.126\pm0.014$	&	$15.5$	&	$6.42\pm0.04$	\\
CrA 047	&	$14.5\pm0.3$&$24.3\pm0.4$	& $16.2\pm0.6$ &	$9.0\pm0.3$  	&	$16.2$	&	$<0.37$     	&	$19.0$	&	$11.73\pm0.11$	\\
CrA 044	&	$12.4\pm0.2$&$20.9\pm0.4$	& $22.2\pm0.8$ &	$7.28\pm0.13$	&	$22.2$	&	$<0.19       $	&	$18.2$	&	$10.63\pm0.09$	\\
CrA 050	&	$10.5\pm0.4$&$7.6\pm0.4$	& $14.9\pm1.1$ &	$2.8\pm0.2$	    &	$14.9$	&	$< 0.38      $	&	$18.0$	&	$5.91\pm0.08$	\\
CrA 151	&	$6.4\pm0.4$	&$5.7\pm0.8$	& $12.5\pm0.3$ &	$9.8\pm0.4$   	&	$12.5$	&	$0.52\pm0.06$	&	$11.2$	&	$2.47\pm0.04$	\\
\hline
\end{tabular}
\tablefoot{\tablefoottext{a}{The column densities in Oph 3 and Oph 4 are summed over the two velocity components; the excitation temperatures are, hence, not given.}}
\end{table*}

\subsection{Uncertainties in the column density estimations\label{sec:unc}}
We made several assumptions in the methodology to compute column densities described in Sect.~\ref{sec:Ncols}. Here, we discuss the main ones.
\subsubsection{Optical depths}
In this work we used \htcop and \cdo to trace the total abundance of the corresponding main isotopologues, in the assumption that they are optically thin. To verify this point, the fitting routine implemented in \textsc{pyspeckit} returns the peak optical depth $\tau$ of each transition. We computed the mean and standard deviation of the $\tau$ distributions, obtaining: $\tau (\rm H^{13}CO^+ \; 2-1) = 0.44 \pm 0.26$,  $\tau (\rm H^{13}CO^+ \; 3-2) = 0.26 \pm 0.12$, and  $\tau (\rm C^{18}O \; 2-1) = 0.73 \pm 0.23$. All these values are consistently lower than $1.0$, indicating general optically-thin conditions. \par 
For completeness, we report also the values for the remaining lines:  $\tau (\rm DCO^+ \; 3-2) = 0.58 \pm 0.99$,  $\tau (\rm DCO^+ \; 4-3) = 0.30 \pm 0.27$, and  $\tau (\rm oH_2D^+ \; 1_{1,0}-1_{1,1}) = 0.22 \pm 0.20$. The mean values are $<1.0$. All the transitions are on average optically thin across the sample. The optical depth values in the whole sample are available in Table ~\ref{tab:fit_params}.

\subsubsection{Isotopic ratios}
\begin{figure}[!h]
    \centering
    \includegraphics[width=\linewidth]{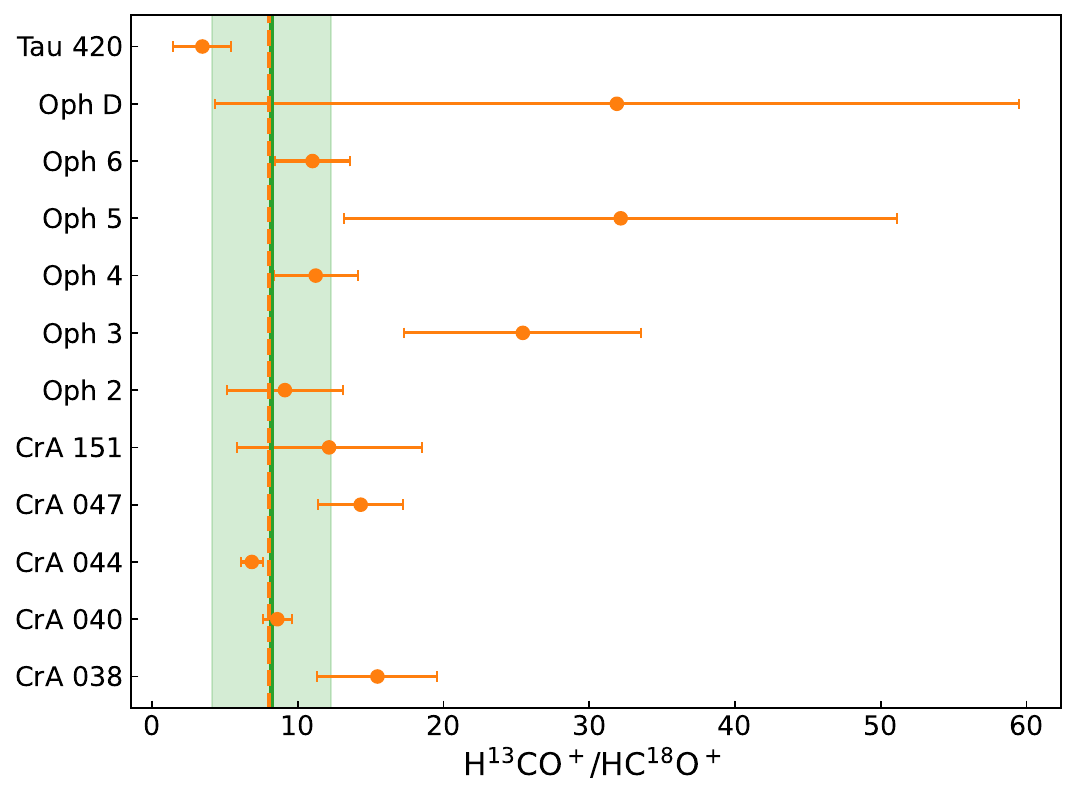}
    \caption{The orange points show the measured flux ratio between the \htcop and $\rm HC^{18}O^+$ $(2-1)$ lines, in those cores where both transitions are detected (labelled on the y-axis). Errorbars show the $3\sigma$ level. The vertical dashed orange line is the weighted average of the sample. The expected value $557/68=8.2$ is shown with the vertical green line, and the shaded green area shows a variation of a factor $2$ around it.  }
    \label{fig:isotopic_ratio}
\end{figure}

When using rarer isotopologues, we assumed the isotopic ratios $\rm ^{16}O/ ^{18}O = 557$ and $\rm ^{12}C/ ^{13}C = 68$ to infer the total column densities of the main species. However, recent chemical modelling work showed that carbon fractionation may lead to significant variations (up to a factor of 2) of the $\rm ^{12}C/ ^{13}C $ ratio \citep{Colzi20, Loison20}. The expected variations of the oxygen isotopic ratio are more modest \citep{Loison19}. For 12 targets in our sample, we have detected the $\rm HC^{18}O^+ \; (2-1)$ line. We have, hence, compared the flux ratio of the \htcop and $\rm HC^{18}O^+$ transitions (with equal $J$) to the expected value $557/68=8.2$. This is appropriate as the two lines have similar spectroscopic constants (see Table~\ref{tab:lines}), and they are optically thin (see previous subsection). We computed integrated intensities over $N_\mathrm{ch}$ channels with $\rm S/N> 3$ signal in the velocity range $2.8-13.5 \,$\kms. The associated uncertainties are estimated via $\sigma_{\rm II} = rms \times \Delta V_{\rm ch} \times \sqrt{N_{\rm ch}}$. The result is presented in Fig.~\ref{fig:isotopic_ratio}. 
\par
The observed \htcop/$\rm HC^{18}O^+$ values are in the range $3.5-32$, with a weighted average of $8.0$, very close to the expected value $557/68=8.2$. The highest values found in the sample are characterised by the highest uncertainties, as they come from sources where the $\rm HC^{18}O^+$ flux has a low $\rm S/N=3-5$. For 50\% of cases, the measured value is consistent with the expected one, and in 83\% of the sample the value is consistent with a variation of a factor of 2. Given the uncertainties, we conclude that the isotopic ratios we assumed in the analysis are robust. 

\subsubsection{Beam filling factor}
When computing column densities, {we assumed a beam filling factor} $\eta_\mathrm{ff}=1$, which means that the molecular emission is more extended than the APEX angular resolution. This needs to be verified, as {column densities will be underestimated if $\eta_\mathrm{ff} < 1$ (beam-dilution).} At the frequencies of interest, the APEX angular resolution spans the range $16-35''$, which corresponds to $1800-3800\,$au and $4000-8800\,$au for the closest and farthest target, respectively. 
\par
Since {we do not have available maps to check the real extension of the molecular emission, we have instead examined available data from similar sources.} For instance, \cite{Keown16} reported multi-pointing detection of \dcop (3-2) in L492 and L694-2 (the latter is also present in our sample), and the emission appears well extended beyond 10000$\,$au. A similar extension is seen also in L1544 \citep{Redaelli19a}. Emission extended over several thousands of au is often reported for \htcop, even though most studies focus on the (1-0) transition (see e.g. B68, \citealt{Maret07}; L1498, \citealt{Maret13}; several cores in the L1495 filament, \citealt{Punanova18}). \cdo is an abundant species, which combined with the low critical density of the (2-1) transition makes its emission likely extended (cf. the SEDIGISM survey that used this line to map the ISM in the inner Galaxy, \citealt{Schuller21}). \par
The case of \ohhdp is more complex, as few maps of this species have been observed in the low-mass regime, to our knowledge. \cite{Vastel06} mapped the \olineh transition in L1544 with the CSO telescope (angular resolution $20''$), and detected it over several beams. The map obtained towards Ophiuchus A SM1N with the James Clark Maxwell Telescope (JCMT) shows emission over $\sim 5000\,$au around the core \citep{Friesen14}. In Ophiuchus HMM1, \cite{Harju24} used APEX/LAsMA to map the \olineh line, which is emitted over more than $10000\,$au. We conclude that it is reasonable to expect this transition to fill the $16''$ APEX beam in the majority of our targets. This is further supported by the fact that for eight cores in the sample, we detected the line not only in the central LAsMA beam, but also in some of the external beams (which are separated by $40''$). \par
{Even though we can safely exclude beam-dilution issues, we stress that the difference in beam size at different frequencies might cause the underestimation of the column density of some tracers with respect to that of \ohhdp, which presents the highest resolution. In the final results this problem is mitigated by the fact that the underestimated column densities appear in Eq.~\ref{eq:crir_observables} as ratios of quantities estimated at similar resolution ($R_\mathrm{D}$ from the \htcop and \dcop lines at 2 and 1$\,$mm, and $X\rm (CO)$ from CO observations at $\sim 30''$ resolution and \textit{Herschel} maps at $37''$). We cannot exclude, however, the underestimation of \crir due to that of $N_\mathrm{col}(\text{\ohhdp})$}.

\subsection{Deuteration levels and  CO depletion factor\label{sec:fd_rd}}
\begin{figure}[!h]
    \centering
    \includegraphics[width=\linewidth]{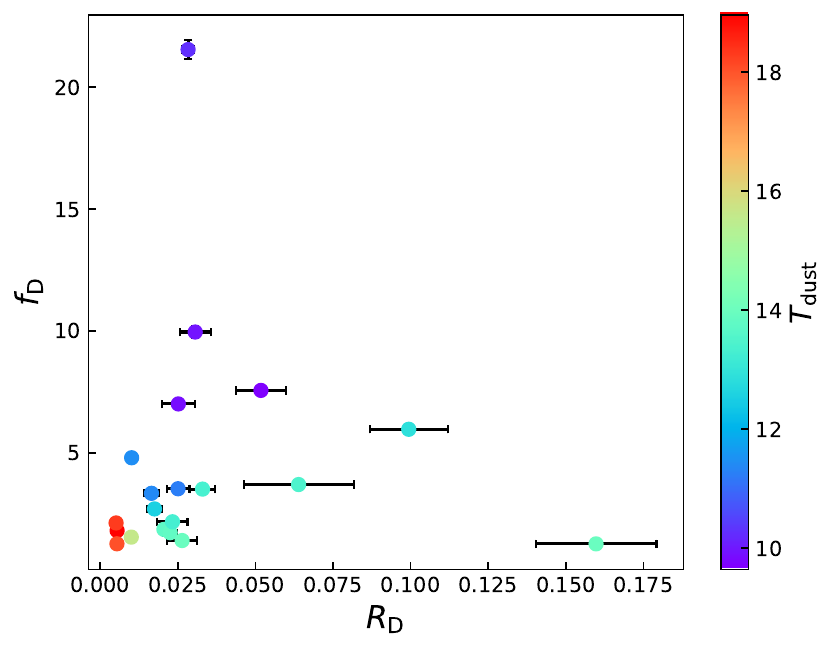}
    \caption{CO depletion factor $f_\mathrm{D}$ as a function of the \hcope deuteration fraction $R_\mathrm{D}$ in the whole sample. The colorscale is determined by the dust temperature value as reported in Table~\ref{tab:cores}.}
    \label{fig:RD_FD}
\end{figure}

We used the column density values obtained in Sect.~\ref{sec:Ncols} to compute the deuteration level and the CO abundance in each target. From the latter, we computed the CO depletion factor $f_\mathrm{D} = X^{\rm st}({\rm CO})/X({\rm CO})$, where $ X^{\rm st}({\rm CO}) = 1.2\times 10^{-4}$ is the standard abundance of undepleted CO {\citep{Lee96,Pineda10,Luo23}}. The values are summarised in Table \ref{tab:CRIR_results}, and their correlation is shown in Fig.~\ref{fig:RD_FD}. \par
The \hcope deuteration factors are in the range $0.5-15$\%, with an average value of $3.4$\%{. Our measured values are} in line with other studies of prestellar and starless cores \citep[cf.][]{Butner95,Tine00}. For a direct comparison of individual targets, {our derived value (2.8\%) in L1544 is consistent with previous result by} \cite{Redaelli19a} {(3.5\%)}; in L183, our value of $5.2 \pm 0.8$\% is in agreement with the range $3-5$\% reported by \cite{Juvela02}. The CO depletion factors {in our sample are in the range $1.3 - 21$}, typical for cold and dense cores, as found by \cite{Bacmann02,  Crapsi05, Lippok13}. \par
Both these quantities are strongly correlated with the core {dust} temperature, as can be seen in Fig.~\ref{fig:RD_FD}, where the colour scale of the data points is determined by the dust temperature value derived with \textit{Herschel}. The lowest deuteration and depletion values are found in cores where $T_\mathrm{dust}>15\,$K, becoming comparable to the CO desorption temperature ($\approx 20\, $K, cf. \citealt{Oberg11}). This immediately lowers the CO depletion factor and{, in this physical conditions, the efficiency of deuteration processes is also lowered \citep{dalgarno84, Ceccarelli14}.} The highest deuteration level is found in Oph 5, which presents an intermediate temperature within the sample ($T_\mathrm{dust}=13.9\,$K). 
However, we highlight that the molecular line emission towards this source presents hints of double-velocity components that cannot be fit for all species (cf. the figure in Appendix~\ref{app:full_figures}). {We highlight that in the sources with the lowest $f_\mathrm{D}$ values, Eq.~\ref{eq:crir_observables} might underestimate the \crir value. This, however, does not change the results of our analysis (see Sect. \ref{sec:disc}).}
\subsection{Estimates of \crir \label{sec:CRIR}}
\begin{figure}[!t]
    \centering
    \includegraphics[width=\linewidth]{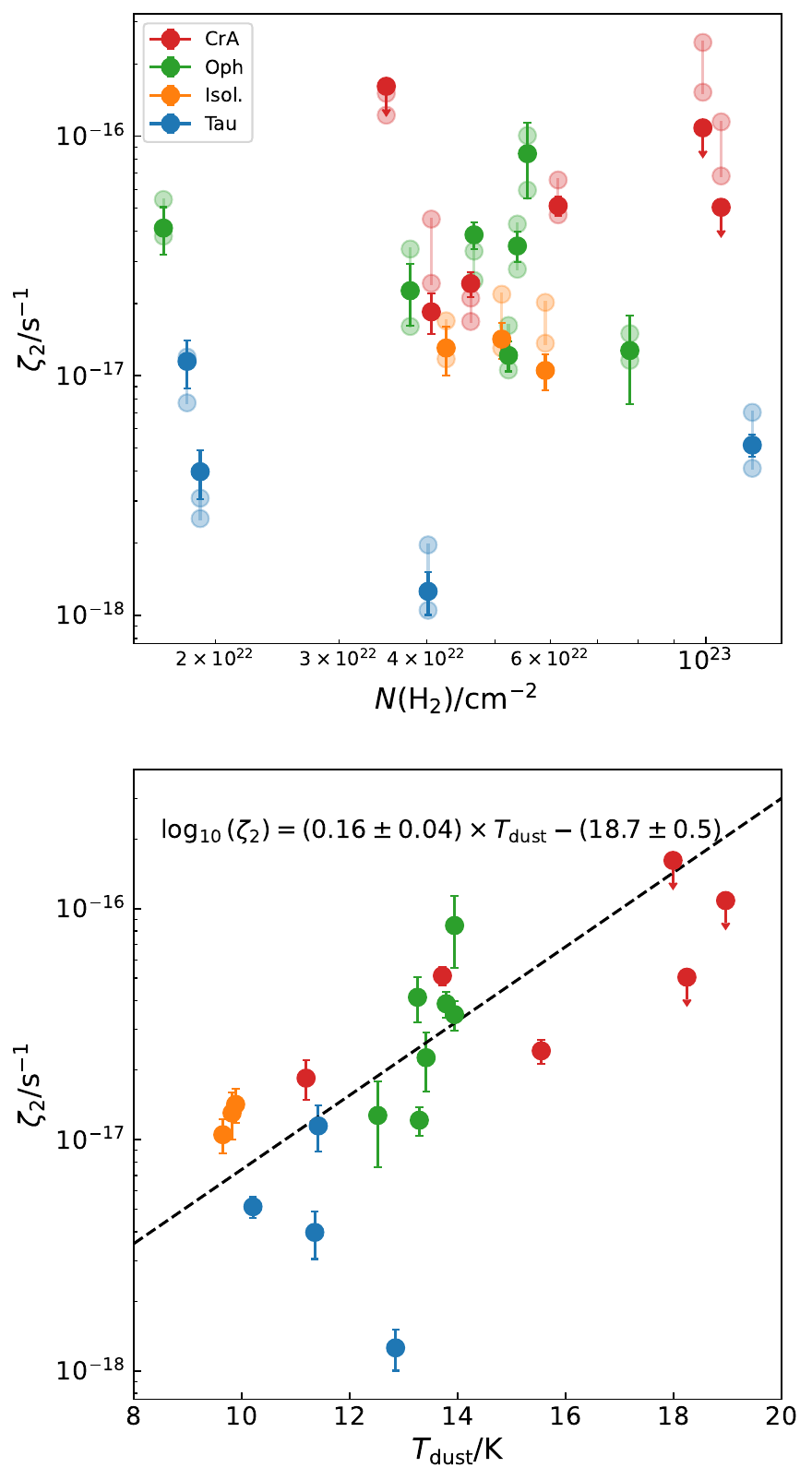}
    \caption{\textit{Upper panel:} The obtained \crir values are plotted as a function of the total gas column density towards each core (including the column density integrated along the line of sight). The solid points and errorbars are computed using $L=0.1\, \rm pc$. The upper limits are shown with downward arrows. The transparent points show the range of \crir values obtained when $L$ is computed using the 40\% or 60\% contours of $N\rm (H_2)$, as described in the Main Text. \textit{Lower panel:} The \crir values (computed for $L=0.1 \, \rm pc$) are plotted as a function of the cores' dust temperature{, as} derived from the \textit{Herschel} maps. The dashed line shows the linear regression computed in the $\log_{10}(\zeta_2)$-\tdust plane, and the computed regression coefficients are shown at the top of the panel. In both panels, the data-points are colour-coded by parental cloud (red: Corona Australis, green: Ophiuchus, blue: Taurus, orange: Isolated sources). }
    \label{fig:CRIR_1}
\end{figure}
\begin{table*}[!h]
\centering
    \renewcommand{\arraystretch}{1.2}

\caption{List of \hcope deuteration fractions, CO depletion factors, and \crir values obtained across the sample. The last column reports the range of \crir values obtained when using different $N \rm (H_2)$ contours to estimate $L$. }
\label{tab:CRIR_results}
\begin{tabular}{ccccc}
\hline \hline
Source  &   $R_\mathrm{D}$  &   $f_\mathrm{D}$  &  \multicolumn{2}{c}{$\zeta_2/ 10^{-17} \, \rm s^{-1}$}\\
        &    $\mathbf{\times 10^{-2}}$  &                   &   $L=0.1\,$pc     &   $L$ from 40-60\% N($\rm H_2$) \\
\hline 
Tau 410	&	$1.01\pm0.16$	&	$4.81\pm0.11$	&	$1.2\pm0.3 $	&	$0.8-1.2$	\\
Tau 420	&	$9.9\pm1.3$ 	&	$5.97\pm0.09$	&	$0.13\pm0.03$	&	$0.1-0.2$	\\
TMC1-C	&	$1.7 \pm0.2 $	&	$3.35\pm0.06$	&	$0.40\pm0.09$	&	$0.25-0.31$	\\
L1544	&	$2.8\pm0.2 $	&	$21.5\pm0.4$	&	$0.51\pm0.06$	&	$0.4-0.7$	\\
L183	&	$5.2\pm0.8 $	&	$7.57\pm0.11$	&	$1.05\pm0.18$	&	$1.4-2.0$	\\
L429	&	$3.1 \pm0.5 $	&	$10.0\pm0.3$	&	$1.4 \pm0.2$	&	$1.3-2.2$	\\
L694-2	&	$2.5\pm0. 5$	&	$7.01\pm0.14$	&	$1.3\pm0.3 $	&	$1.2-1.7$	\\
Oph 1	&	$6.4 \pm1.8 $	&	$3.71\pm0.08$	&	$2.3 \pm0.7$	&	$1.6-3.4$	\\
Oph 2	&	$3.3 \pm0.4$	&	$3.52\pm0.03$	&	$1.21\pm0.17$	&	$1.1-1.6$	\\
Oph 3	&	$1.8 \pm0.2 $	&	$2.71\pm0.09$	&	$1.3\pm0.7 $	&	$1.2-1.5$	\\
Oph 4	&	$2.6\pm0.5 $	&	$1.41\pm0.09$	&	$8 \pm 4$	    &	$5.9-10.1$	\\
Oph 5	&	$16\pm2    $	&	$1.274\pm0.012$	&	$3.5 \pm0.5$	&	$2.8-4.3$	\\
Oph 6	&	$2.3 \pm0.2$	&	$1.732\pm0.013$	&	$3.9 \pm0.5$	&	$2.5-3.3$	\\
Oph D	&	$2.3\pm0.5 $	&	$2.18\pm0.04$	&	$4.1\pm0.9$ 	&	$3.8-5.4$	\\
CrA 038	&	$2.05\pm0.15$	&	$1.87\pm0.03$	&	$5.1\pm0.5 $	&	$4.7-6.6$	\\
CrA 040	&	$1.00\pm0.05$	&	$1.550\pm0.010$	&	$2.4\pm0.3 $	&	$1.7-2.1$	\\
CrA 044	&	$0.511\pm0.014$	&	$2.13\pm0.02$	&	$< 5.05$	&	$<6.8-11.5$	\\
CrA 047	&	$0.544\pm0.019$	&	$1.82\pm0.02$	&	$< 10.8  $	&	$<15.3-24.6$	\\
CrA 050	&	$0.54\pm0.05$	&	$1.28\pm0.02$	&	$< 16.2 $	&	$<12.2-15.1$	\\
CrA 151	&	$2.5\pm0.4 $	&	$3.53\pm0.06$	&	$1.9 \pm0.4$	&	$2.4-4.5$	\\

\hline
\end{tabular}
\end{table*}

Using the $R_\mathrm{D}$ and $X\mathrm{(CO)}$ values, we applied Eq.~\eqref{eq:crir_observables} to infer the cosmic-ray ionisation rate values. We initially assumed a value of $L=0.1\, \rm$pc for the source size, which is often regarded as a typical size of low-mass cores in the local ISM \citep[cf.][]{Andre00, Pineda23}. A further discussion on the choice of this parameter is presented later. The obtained values and associated uncertainties are reported in the fourth column of Table~\ref{tab:CRIR_results} and in Fig.~\ref{fig:CRIR_1}, as a function of the total gas column density $N\rm (H_2)$ (upper panel). In the three sources (CrA 044, 047, and 050) where \ohhdp is not detected, we report $3\sigma$ upper limits {of \crir}. The \crir values are scattered over almost two orders of magnitude, ranging from a minimum of $1.3 \times 10^{-18} \, \rm s^{-1}$ to a maximum of $8.5 \times 10^{-17} \, \rm s^{-1}$ (excluding the upper limits). The scatter is significantly larger than the intrinsic uncertainty of the method \citep[estimated as a factor of $2-3$, see][]{Redaelli24}. We discuss possible explanations for the observed scatter in Sect.~\ref{sec:disc}.
\par

The choice of the $L$ value is crucial, as the analytic expression for \crir is inversely proportional to it. The topic has been discussed lengthy in \cite{Bovino20} and \cite{Redaelli24}. A possible approach is to estimate $L$ from the maps of total column density presented in Sect.~\ref{sec:H2_tdust}, using the equivalent radius computed from a given contour of $N\rm (H_2)$. Choosing which contour is, however, not straightforward. The cores present different contrast levels, i.e. the ratio between the peak $N\rm (H_2)$ value at the core's centre and the surrounding average $N\rm (H_2)$ varies significantly, from a factor of a few to more than one order of magnitude. Furthermore, several cores are embedded in crowded environments (in particular, in Ophiuchus and Corona). The selection of closed $N\rm (H_2)$ contours is, hence, highly dependent on the cores' parental environments. To take into account the uncertainty associated with the $L$ value, we have computed the equivalent diameter corresponding to 40\% and 60\% of the $N\rm (H_2)$ peak at the cores' position. {These levels lead to close contours in the whole sample}. These values are presented in Table~\ref{tab:cores}. We then computed a range of corresponding \crir, using the range of $L$ values in Eq.~\ref{eq:crir_observables}, and we report them in the last column of Table~\ref{tab:CRIR_results} and in the upper panel of Fig.~\ref{fig:CRIR_1} using shaded points. For most sources, the \crir value obtained using $L=0.1\, \rm pc$ is consistent, within uncertainties, with the range of values computed adopting a variable $L$, with only three exceptions (CrA 151, L183, and Oph 6)\footnote{Here, we do not take into considerations the three sources with upper limits on \ohhdp.}. \par
{A second possibility to infer $L$ is to use the detected lines to trace the total gas volume density and, thus, $L = N\mathrm{(H_2)}/n\mathrm{(H_2)}$. In Appendix~\ref{app:RADEX} we perform LVG modelling of the \htcop transitions, which are available in the majority of the sample. The results show that even if it is true that in several targets the  $n\mathrm{(H_2)}$ values derived from $L=0.1\rm \, pc$ are underestimated, this underestimation is {in line with the method's error}, and this does not affect the main conclusions of this work. We conclude that the choice of $L$ will not significantly impact our analysis and results.}

\subsection{Ionisation degree and dynamical state of the cores \label{sec:xe}}
\begin{figure}[!h]
    \centering
    \includegraphics[width=1.\linewidth]{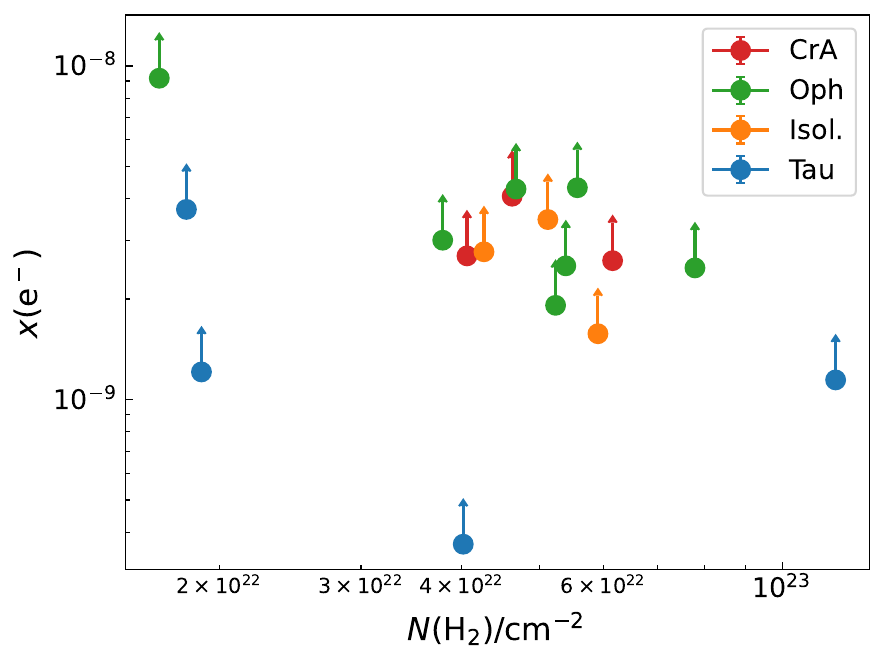}
    \caption{Estimation of the ionisation fraction \xe using Eq.~\eqref{eq:xe}, as a function of the cores' central total column densities, in the targets with detected \ohhdp emission. We neglected the contribution of $\rm N_2H^+$ and $\rm N_2D^+$. As in Fig.~\ref{fig:CRIR_1}, the data-points are color-coded according to the parental cloud.}
    \label{fig:xe}
\end{figure}

The ionisation degree of the gas is measured by the electron abundance, or electron fraction \xe. In the assumption that free electrons are donated by neutral species, and that the charge of dust grains is negligible, one can compute \xe by assuming the gas charge balance. {\cite{Latrille25}} introduced the following proxy for \xe:
\begin{equation}
 \text{\xe} = X (\mathrm{H_3^+} )  + X (\mathrm{\text{\hcope}} )+ X (\mathrm{\text{\dcop}} ) + X (\mathrm{N_2H^+}  )+ X \mathrm{(N_2D^+)}  \, ,
 \label{eq:xe}
\end{equation}
where $ X (\mathrm{H_3^+} )= X (\mathrm{oH_2D^+} )/3 R_\mathrm{D} $. The authors tested it on a set of magneto-hydrodynamic simulations, finding that it reproduces the true \xe values with a scatter of $0.5\,$dex. According to the chemical model of those authors, the main contributions to Eq.~\eqref{eq:xe} come from $\rm H_3^+$ and \hcope. We used this equation to measure \xe in each core of the sample, neglecting the diazenylium terms for which we have no estimate, considering only sources with the detection of \ohhdp. The obtained values are lower limits to the true gas ionisation degree. Figure~\ref{fig:xe} shows the resulting plot of \xe as a function of the cores' column density. \par
The obtained values are found in the range $10^{-9} - 10^{-8}$, with the majority of cores showing an ionisation degree of $2-4 \times 10^{-9}$. These values are in line with literature results in similar sources \citep{Caselli02b, Maret07}, but they are substantially lower than those derived in NGC1333 by \cite{Pineda24}.
\par

The ionisation degree is directly proportional to the timescale for ambipolar diffusion $t_\mathrm{AD}$, i.e. the process of drift between the neutral species and the ionised one due to imperfect coupling between the two gas flows. In particular, $t_\mathrm{AD}$ can be computed as \citep{Spitzer78, Shu87}:

\begin{equation}
t_\mathrm{AD} = 2.5 \times 10^{13}  \times \text{\xe} \rm \, yr \; .
\end{equation}
We have used the computed values of \xe to estimate $t_\mathrm{AD}$ in each core. These values are also lower limits. We compare them to the free-fall timescales ($t_\mathrm{ff}$), to assess the dynamical stability of the cores:
\begin{equation}
    t_\mathrm{ff} = \sqrt{\frac{3 \pi}{32 \, G \mu m_\mathrm{H} n(\rm H_2) }} \; . 
\end{equation}
We adopt $\mu = 2.33$ for the mean molecular weight of the gas and $n({\rm H_2}) = N(\rm H_2)/0.1 \, \rm pc$, using the $N(\rm H_2)$ tabulated values reported in Table ~\ref{tab:cores}. { We stress that this approach is underestimating the true central densities of the cores, because it involves a column density value that is averaged on the \textit{Herschel} beam size and it assumes that the volume density is constant along the line of sight. However, in Appendix~\ref{app:RADEX}} we show that the underestimation is likely of a factor of $\sim 2$, and it does not affect the whole sample. \par
\begin{figure}[!h]
    \centering
    \includegraphics[width=1.\linewidth]{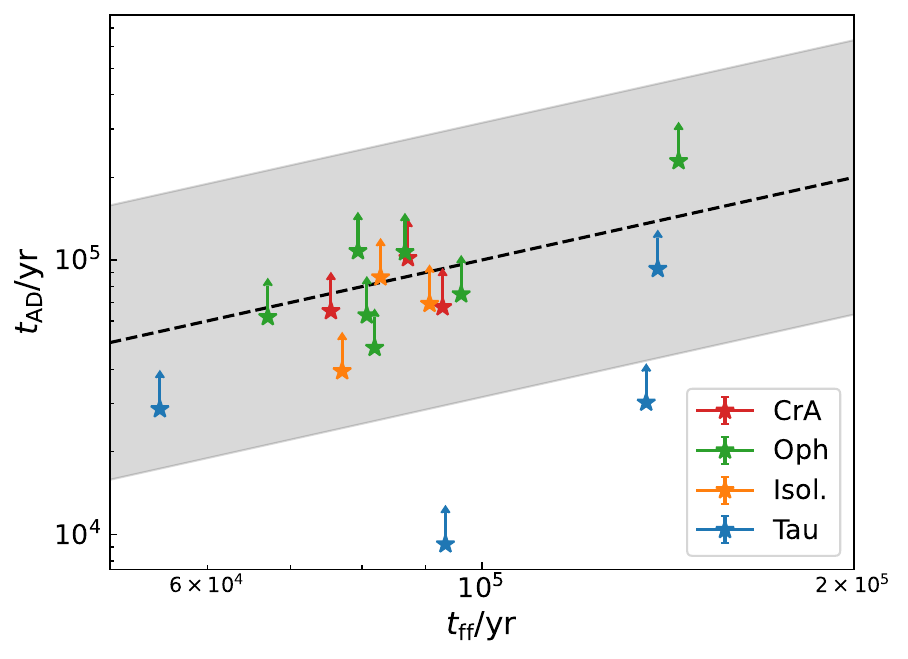}
    \caption{Scatterplot of the ambipolar diffusion timescale $t_\mathrm{AD}$ as a function of the free-fall time $t_\mathrm{ff}$ in the targets with detected \ohhdp emission. The $t_\mathrm{AD}$ values are lower limits as \xe estimated from \eqref{eq:xe} represents a lower limit. The datapoints are color-coded by environment, as in Fig.~\ref{fig:CRIR_1}. The dashed line is the 1:1 relation, whilst the shaded area is a 0.5dex scatter around it (corresponding to the deviation of Eq.~\ref{eq:xe} from the true value estimated by {\cite{Latrille25}}.  }
    \label{fig:tad}
\end{figure}

Figure~\ref{fig:tad} compares the ambipolar diffusion and free-fall timescales in all the sources. Considering the limitations of the method, we can conclude that the derived $t_\mathrm{AD}$ values are on average consistent with, or larger than, the corresponding $t_\mathrm{ff}$. For the Oph 1-6 cores, this is in line with \cite{Bovino21}, who found that these sources are consistent with model of fast collapse on shorter timescales than the ambipolar diffusion ones.

\section{Discussion\label{sec:disc}}
\subsection{The role of the environment}
The resulting \crir values shown in Fig.~\ref{fig:CRIR_1} are coloured by parental cloud. A trend can be seen: cores in Ophiuchus and Corona Australis present on average higher ionisation rates than in Taurus or in the more isolated cores (L183, L694-2, and L429, which is found close to the Aquila Rift). The weighted averages in each environment (excluding upper limits) are $\langle \zeta_2 \rangle = (2.8 \pm 0.2) \times 10^{-17} \, \rm s^{-1}$ in Corona, $(1.76 \pm 0.14) \times 10^{-17} \,\rm s^{-1}$ in Ophiuchus, $(0.23\pm 0.03)\times 10^{-17} \, \rm s^{-1}$ in Taurus, and $(1.22 \pm 0.12)\times 10^{-17} \, \rm s^{-1}$ for the three more isolated sources. The differences, in particular between Taurus and Ophiuchus or Corona, are significant considering both the uncertainties reported in Table~\ref{tab:CRIR_results} and the intrinsic error of the method. This points toward the role of the environmental properties in setting the average ionisation rate at the core scales. A similar conclusion was also reached by \cite{Sabatini23} in high-mass star-forming regions, where the authors highlighted that cores embedded in the same parental clump showed similar \crir values, whilst significant differences were found when comparing distinct clumps. \par
We have investigated which environmental physical parameters have the strongest influence on the overall ionisation rate of the cores. From the upper panel of Fig.~\ref{fig:CRIR_1}, {it appears that the differences in column density among these cores may not be the primary factor contributing to the variations in \crir}. There is no clear correlation between \crir and $N \rm (H_2)$, and the cores embedded in different clouds cover similar ranges of $N \rm (H_2)$ values. On the other hand, the regions analysed differ by dust temperature. Looking at the values in Table~\ref{tab:cores}, the cores' temperatures are $11-19\,$K in Corona, $\sim 13.5 \, \rm K$ in Ophiuchus, $10-13 \, \rm K$ in Taurus and $< 10 \,$K in the three isolated sources. The trend is shown in the lower panel of Fig.~\ref{fig:CRIR_1}, where we show also the linear regression fit to the values (upper limits excluded) in the semi-logarithmic plane. The Spearman correlation coefficient is $0.65$, indicating a high probability of positive correlation between the two quantities. The associated $p-$value is $0.004 \ll 0.05$, hence rejecting the null-hypothesis that the two quantities are not correlated. {In Appendix~\ref{app:RADEX} we show that the correlation remains valid when the gas volume density and the $L$ parameter are inferred from the molecular data, instead of the \textit{Herschel} data.}
\par
The different temperatures can be linked to the star-formation activity in the regions. Five of the six cores selected in Corona are found in the proximity of the so-called Coronet cluster, which hosts several YSOs at distinct evolutionary stages (cf. \citealt{Chini03, Sicilia-Aguilar13}). For CrA 044, 047, and 050 (the latter a known protostellar core) we report only upper limits on \ohhdp and \crir. This is because the higher temperatures caused by the nearby protostellar activity favour the CO desorption from the dust grains back to the gas phase, as shown by the low values of CO depletion we measured ($f_\mathrm{D} = 1-2$). CO, in turn, quickly destroys \hhdp. The lowest \crir value in the Corona cores is found in CrA 151, which is located $\sim 5\, \rm pc$ away from the Coronet cluster region and is a quiescent core with high levels of deuteration {\citep{Redaelli25}}. Ophiuchus is also a very active star-forming region, with approximately $\sim 300$ YSO sources identified \citep{Dunham15}. On the contrary, the cores in Taurus and the isolated sources are found in relatively quieter environments, as proved by both the low temperature values and the low {line velocity dispersion}s found in these cores (cf. Table~\ref{tab:fit_params}). The main exception is L429, close to the active region of the Aquila Rift. In further support to the scenario just described, L429 presents the highest \crir value in this subset, even though the difference is within the uncertainties.  \par
The correlation between the star-formation activity and the ionisation rate of the gas is supported by recent theoretical works, which showed that low-mass protostars can be a source of local re-acceleration of cosmic rays \citep{Padovani15, Padovani16}, either at the protostellar surface or in the shocks along the protostellar jets. This can explain the detection of non-thermal radiation in these jets \citep[cf.][]{Ainsworth14}. Regions with strong protostellar activity, therefore, might be characterised by higher levels of ionisation due to CRs, as first noted by \cite{Ceccarelli14b} and seen also in the maps of \cite{Pineda24}. Based on YSOs count, \cite{Heiderman10} and \cite{Evans14} estimated the surface density of star formation $\Sigma (SFR)$ in several Gould Belt Clouds, and reported values of $2.44 \, \rm M_\odot \, yr^{-1} \, kpc^{-2} $ in Ophiuchus, 3.47 in Corona, and 0.15 in Taurus, respectively, which further supports the scenario of correlation between star formation activity and average ionisation level. 

\subsection{Comparison with current models of CR propagation}
\begin{figure}[!h]
   \centering
    \includegraphics[width=1.0\linewidth]{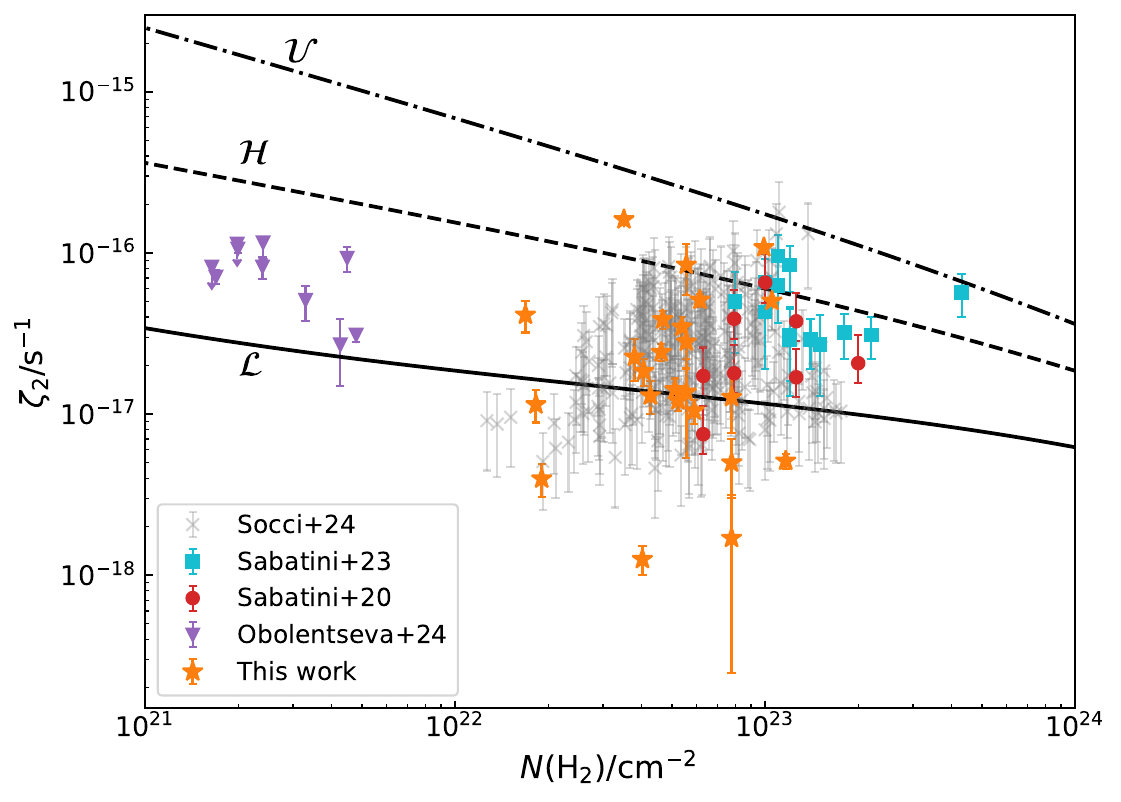}
    \caption{\crir as a function of $N\rm (H_2)$. The theoretical models $\mathscr{L}$ (solid black line), $\mathscr{H}$ (dashed black line), and $\mathscr{U}$ (dotted-dashed black line) are taken from \cite{padovani22}. The new observational estimates obtained in this work are shown with orange stars. The most recent values in the diffuse gas are shown with purple triangles \citep{Obolentseva24}. We also show literature values in dense gas obtained with the method of \cite{Bovino20}, represented with red dots \citep{Sabatini20}, cyan squares \citep{Sabatini23}, and shaded grey crosses \citep{Socci24}.}
    \label{fig:CRIR_mode}
\end{figure}

In this Section, we discuss the obtained results in the context of available measurements of the CR ionisation rate in the Milky Way and in comparison with the prediction of theoretical models of CR propagation and attenuation. Figure~\ref{fig:CRIR_mode} shows a summary of observational estimates taken from literature, as a function of the sources $N\rm (H_2)$, {with a particular focus on those obtained using the same method we implemented}. The only exception {is} the values at low densities ($N \rm (H_2) \lesssim 10^{21}\, \rm cm^{-2}$), which refer to diffuse clouds \citep{Obolentseva24}. These estimate have been re-evaluated \crir using 3D visual extinction maps to infer the volume densities of the sources{. As} a result, they are on average a factor of $9$ lower than previous estimates \citep{Indriolo07, Indriolo12}. The new values obtained in this work are shown with orange stars.
\par
The plots also show the expected trends {for the total ionisation rate computed assuming different models of the energy spectrum of protons} from \cite{padovani22}: model low ($\mathscr{L}$), high ($\mathscr{H}$), and the upper-limit model $\mathscr{U}$ that adopts a steep spectral slope $\alpha = -1.2$. The high model was introduced initially with the scope of explaining the high \crir values obtained in diffuse clouds from $\rm H_3^+$ measurements, and it might not be needed anymore given the results of \cite{Obolentseva24}. As discussed in the previous subsection, we do not see a significant decreasing trend with $N\rm (H_2)$, as the correlation with the environmental temperature {(measured by means of \tdust)} is stronger. We also highlight, however, {that} we explored a relatively small range {of} column density (less than one order of magnitude). Most of our values are found across the $\mathscr{L}$ model, which is the one in agreement with the Voyager data.  Nevertheless, we obtained a significant scatter, both towards low and towards high values. As noted before, we hypothesise that environmental effects are the main driver for it. This strengthens the idea that a universal \crir value cannot be adopted even for sources with similar properties and evolutionary stage. 


\section{Summary and Conclusions\label{sec:summary}}
In this work, we estimated the CR ionisation rate \crir in a sample of dense and cold cores embedded in different star-forming regions in the Solar neighbourhood. We employed the analytical method of \cite{Bovino20}, further corroborated and updated by \cite{Redaelli24}, which correlates \crir with the column density of \ohhdp, the CO abundance, and the deuteration fraction of \hcope. To estimate these quantities, we analysed new, high-sensitivity spectroscopic data observed with the APEX single-dish telescope. Furthermore, we used the column density and dust temperature maps obtained from the analysis of the \textit{Herschel} data to characterise the cores' average $N\rm (H_2)$ and \tdust values. \par
We performed a spectral fit to the observed transitions to infer the molecular column density, assuming C-TEX conditions. The estimated deuteration fractions are in the range $0.5-15$\% and the CO depletion factors span the range $f_\mathrm{D} = 1.3-21$, which are in agreement with expectations from chemical considerations. Both these parameters correlate with the cores' temperature. As \tdust increases and approaches the CO desorption temperature ($\sim 20 \,\rm K$), the CO molecules return to the gas-phase (decreasing $f_\mathrm{D}$) and quickly destroy \ohhdp and, in general, deuterated species, lowering in turn $R_\mathrm{D}$. 
\par
We infer the \crir value in 17 cores. For three cores, \ohhdp was undetected at the sensitivity level of our observations, and we estimated upper limits on \crir. To our knowledge, this represents the largest sample of \crir measurements in low-mass dense cores obtained with a consistent methodology after the one presented by \cite{Caselli98}. The \crir measurements span a range of almost two orders of magnitude, from $1.3 \times 10^{-18}$ to $8.5 \times 10^{-17} \, \rm s^{-1}$. We do not find a significant correlation with the sources' total column density. However, we do find significant differences when comparing cores embedded in distinct environments. We detect a significant positive correlation with the cores' temperature. We speculate that warmer environments are associated with stronger star-formation activity and they {are} more affected by protostellar feedback, which causes the temperature rise. This is consistent with theoretical predictions on how low-mass protostars can be local sources of re-acceleration of cosmic rays. Our work strongly suggests that the gas ionisation state (which is crucial for several physical and chemical processes) is not a common property even in sources of similar kind and evolutionary stage, and, instead, it varies strongly both due to local and environmental effects. Future efforts dedicated to increasing the sample size to more environments will allow a better understanding of which other environmental properties are important to set the ionisation rate in dense gas.

\begin{acknowledgements}
The authors gracefully acknowledge Dr. Jorma Harju for his help and support.
   This publication is based on data acquired with the Atacama Pathfinder Experiment (APEX). APEX has been a collaboration between the Max-Planck-Institut fur Radioastronomie, the European Southern Observatory, and the Onsala Space Observatory.
    This research has made use of data from the Herschel Gould Belt survey (HGBS) project (http://gouldbelt-herschel.cea.fr). The HGBS is a Herschel Key Programme jointly carried out by SPIRE Specialist Astronomy Group 3 (SAG 3), scientists of several institutes in the PACS Consortium (CEA Saclay, INAF-IFSI Rome and INAF-Arcetri, KU Leuven, MPIA Heidelberg), and scientists of the Herschel Science Center (HSC). SB acknowledges BASAL Centro de Astrofisica y Tecnologias Afines (CATA), project number AFB-17002. 
    {GS acknowledges the project PRIN-MUR 2020 MUR BEYOND-2p (``Astrochemistry beyond the second period elements’’, Prot. 2020AFB3FX), the PRIN MUR 2022 FOSSILS (``Chemical origins: linking the fossil composition of the Solar System with the chemistry of protoplanetary disks’’, Prot. 2022JC2Y93), the project ASI-Astrobiologia 2023 MIGLIORA (``Modeling Chemical Complexity’’, F83C23000800005), the INAF-GO 2023 fundings PROTOSKA (``Exploiting ALMA data to study planet forming disks: preparing the advent of SKA’’, C13C23000770005) and the INAF-Minigrant 2023 TRIESTE (“TRacing the chemIcal hEritage of our originS: from proTostars to planEts”; PI: G. Sabatini). GL gratefully acknowledges the financial support of ANID-Subdirección de Capital Humano/Magíster Nacional/2024-22241873 and Millenium Nucleus NCN23{\_}002 (TITANs). MP acknowledges the INAF-Minigrant 2024 ENERGIA ("ExploriNg low-Energy cosmic Rays throuGh theoretical InvestigAtions at INAF"; PI: M. Padovani). {The authors thank the anonymous referee for their comments, which helped improved the quality of the manuscript. }}
    
\end{acknowledgements}

\bibliographystyle{aa}
\bibliography{Literature.bib}

\appendix

\section{Spectral fit in the full sample\label{app:full_figures}}
\begin{figure}[!h]
    \centering
    \includegraphics[width=\linewidth]{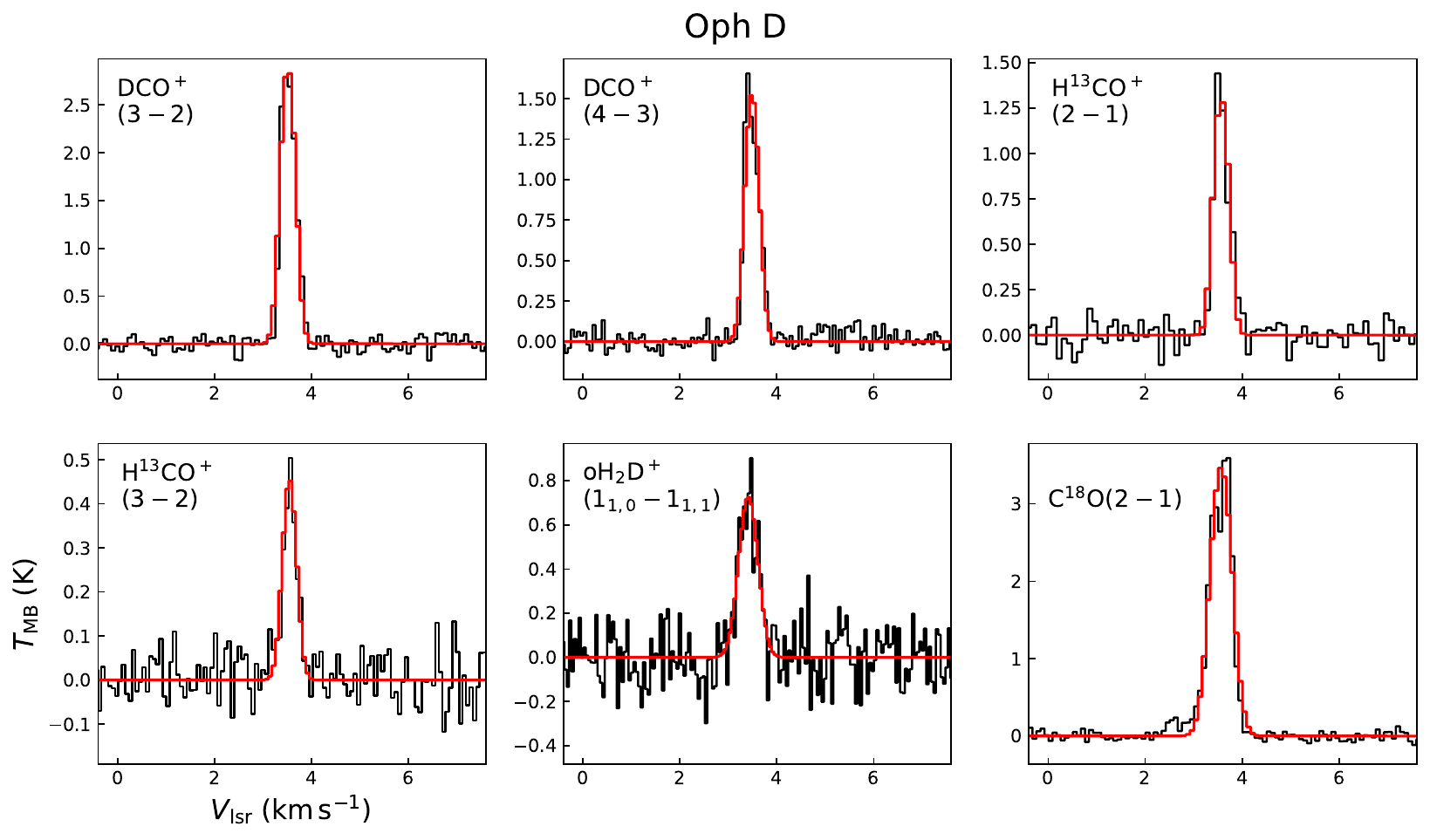}
    \caption{ \label{fig:first_app}}
\end{figure}

\begin{figure}[!h]
    \centering
    \includegraphics[width=\linewidth]{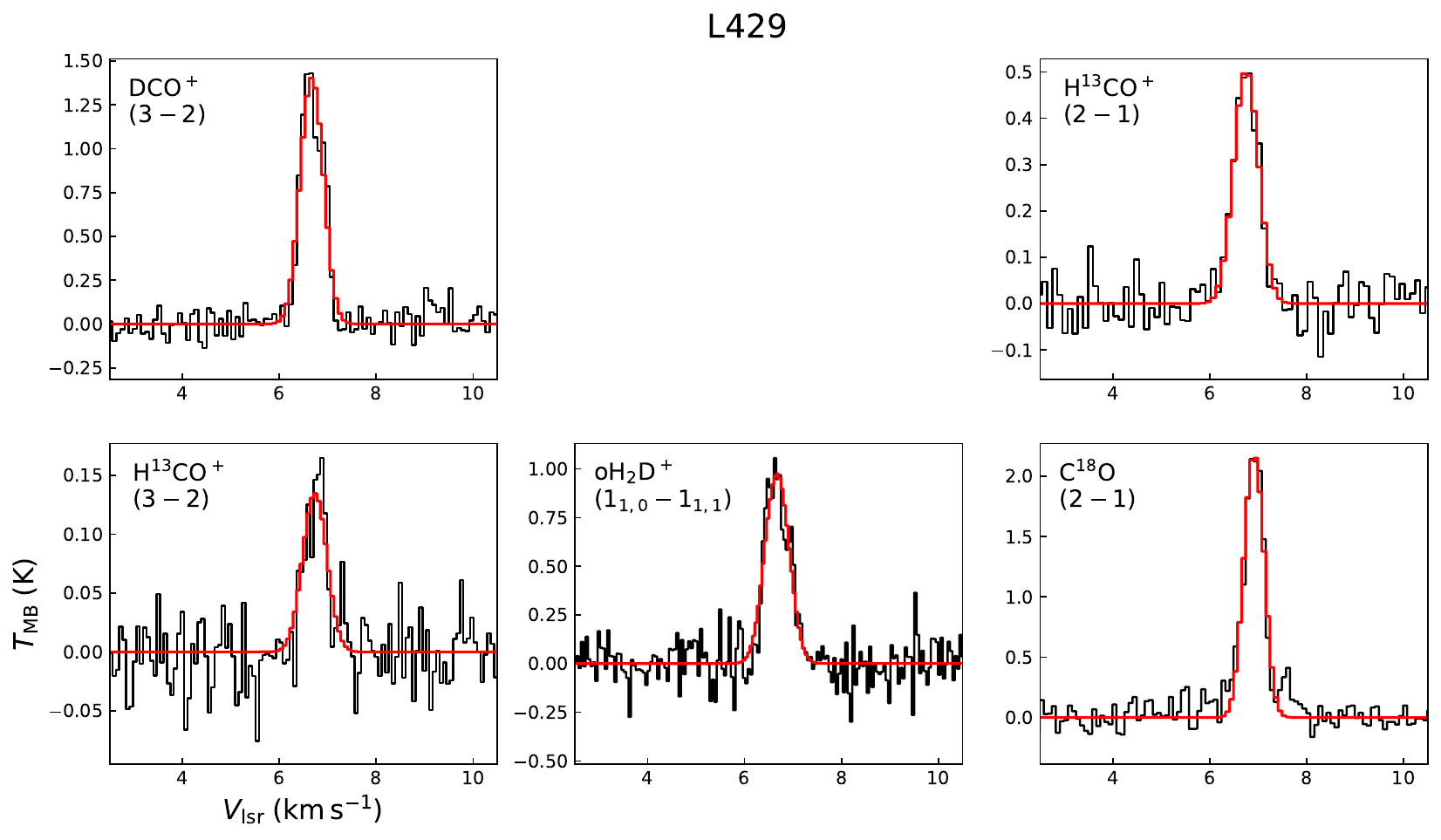}
    \caption{Same as Fig.~\ref{fig:first_app}, but for L429.\label{fig:last_app}}
\end{figure}

\begin{figure}
    \centering
    \includegraphics[width=0.7\linewidth]{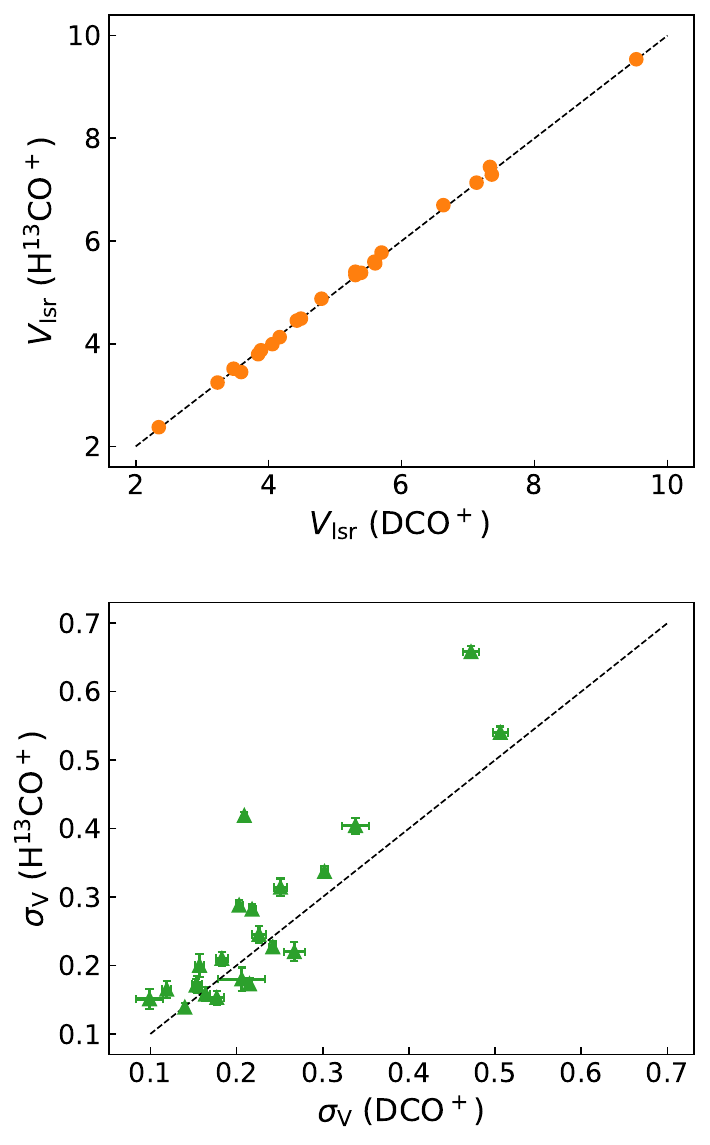}
    \caption{ Comparison of the \dcop and \htcop centroid velocities (top panel) and velocity dispersions (bottom panel). The dashed lines show the 1:1 relation.}
    \label{fig:hcop_dcop}
\end{figure}

Figures \ref{fig:first_app} to \ref{fig:last_app} show the spectral fit performed with \textsc{pyspeckit} in each of the cores not shown in the Main Text. Table ~\ref{tab:fit_params} summarises the best-fit values obtained for the kinematic
parameters ($V_\mathrm{lsr}$ and $\sigma_V$), together with the peak optical depth of each transition. {Figure~\ref{fig:hcop_dcop} shows the comparison of the centroid velocity and dispersion values for \htcop and \dcop.}
\begin{figure}[!h]
    \centering
    \includegraphics[width=\linewidth]{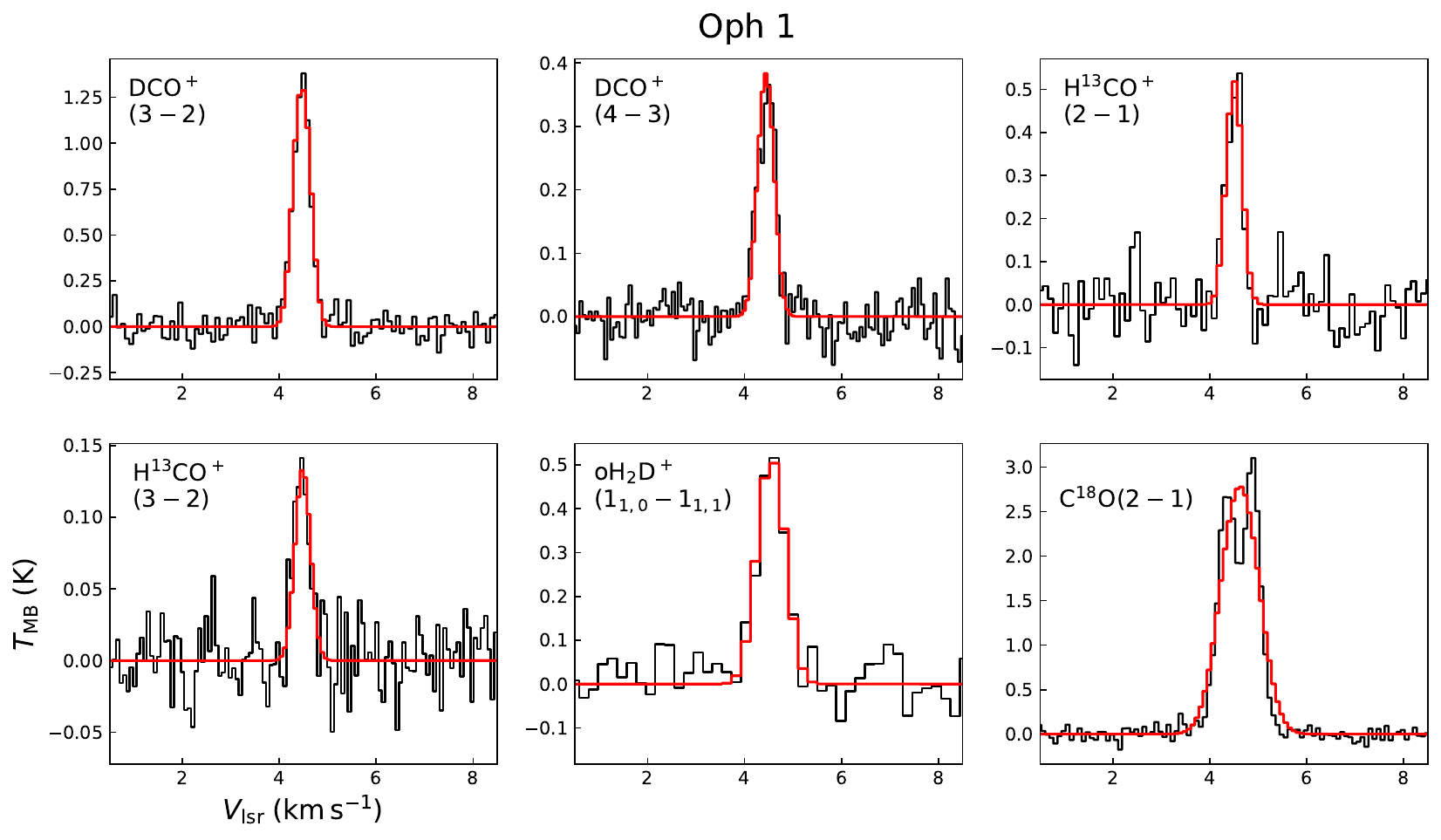}
    \caption{Same as Fig.~\ref{fig:first_app}, but for Oph 1.}
\end{figure}
\begin{figure}[!h]
    \centering
    \includegraphics[width=\linewidth]{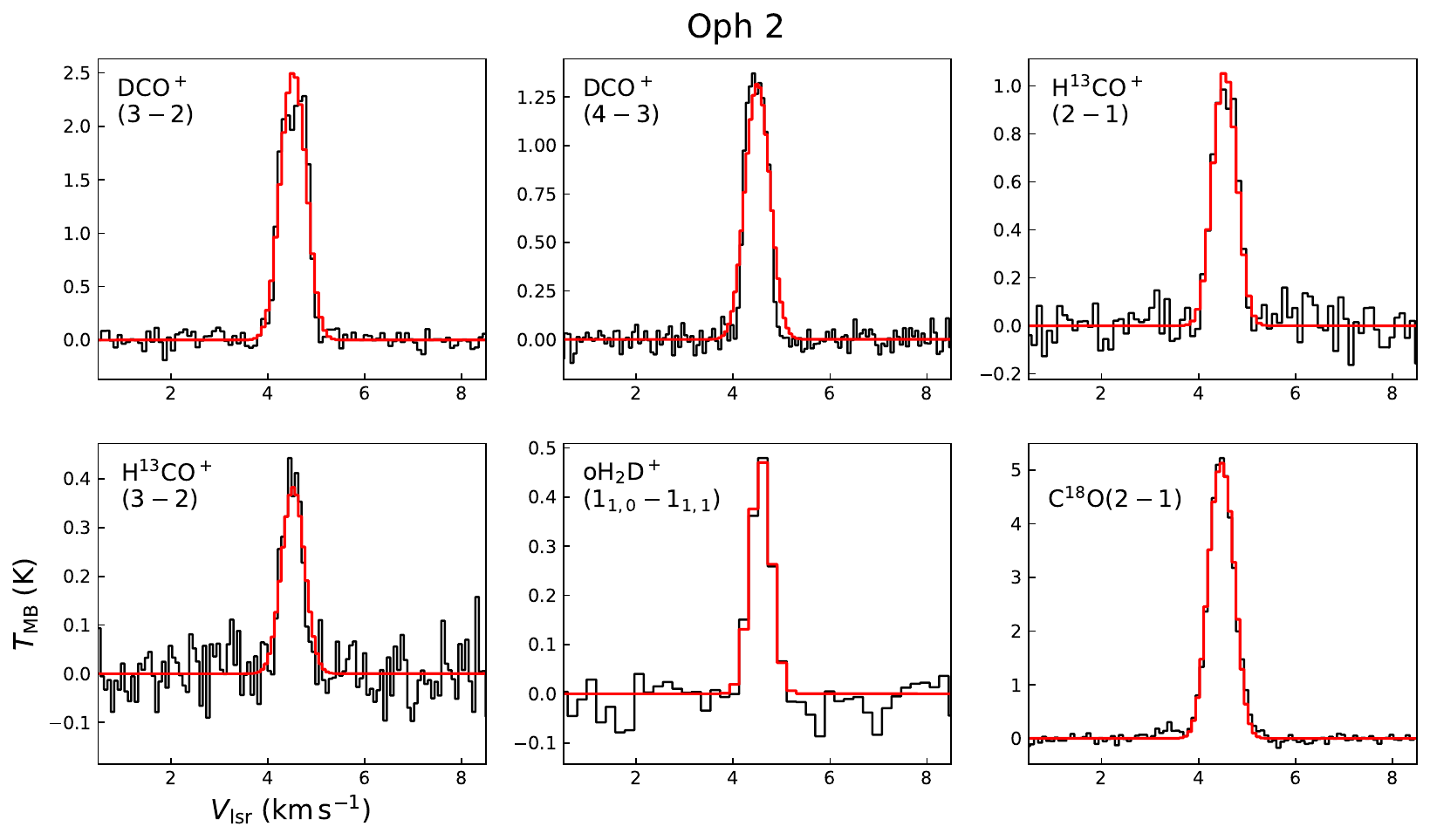}
    \caption{Same as Fig.~\ref{fig:first_app}, but for Oph 2.}
\end{figure}
\begin{figure}[!h]
    \centering
    \includegraphics[width=\linewidth]{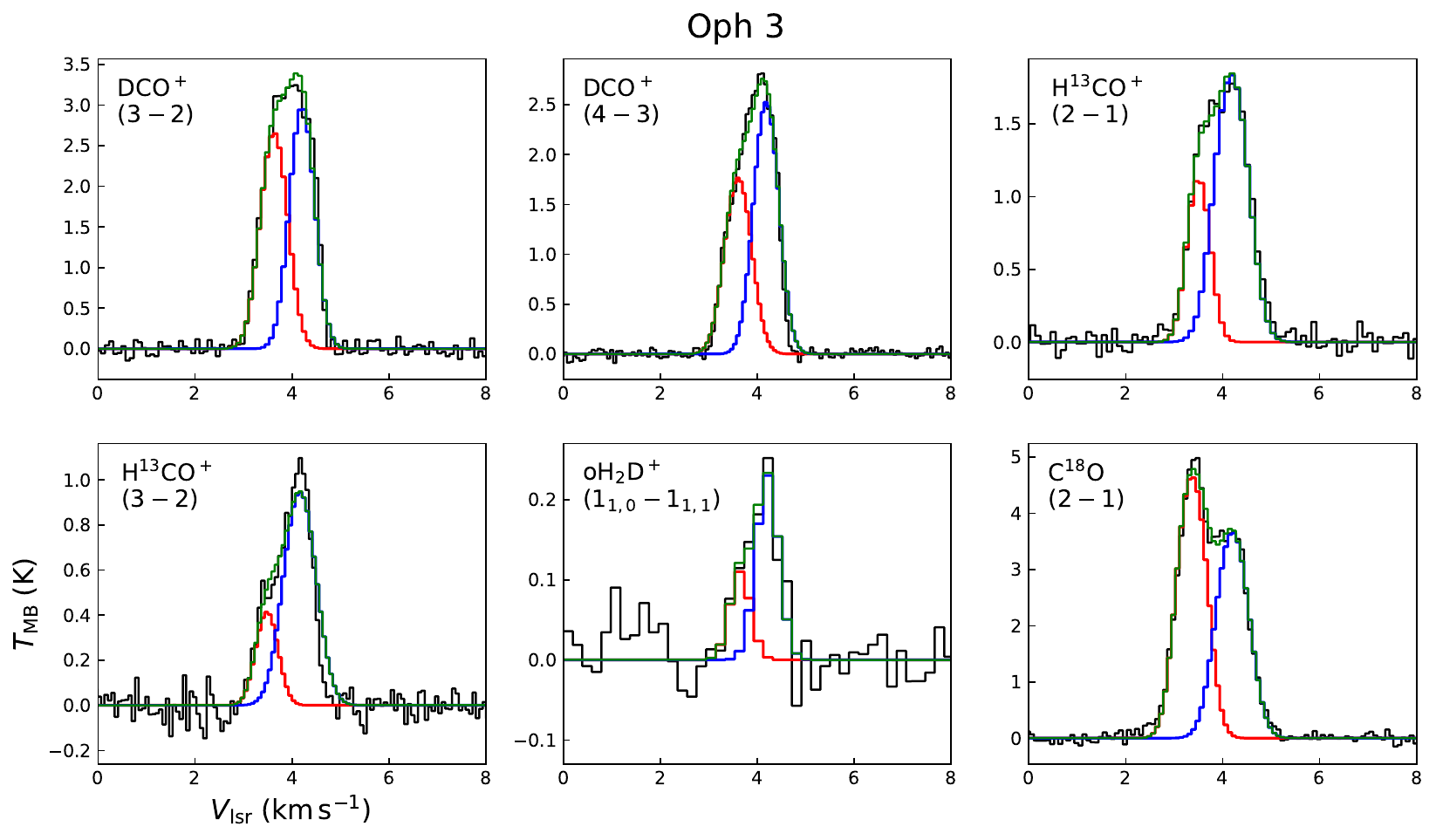}
    \caption{Same as Fig.~\ref{fig:first_app}, but for Oph 3. As for Oph 4, we performed a two-velocity-component fit. The two distinct components are shown with the red and blue curves, whilst the green curve shows the total fit.}
\end{figure}

\begin{figure}[!h]
    \centering
    \includegraphics[width=\linewidth]{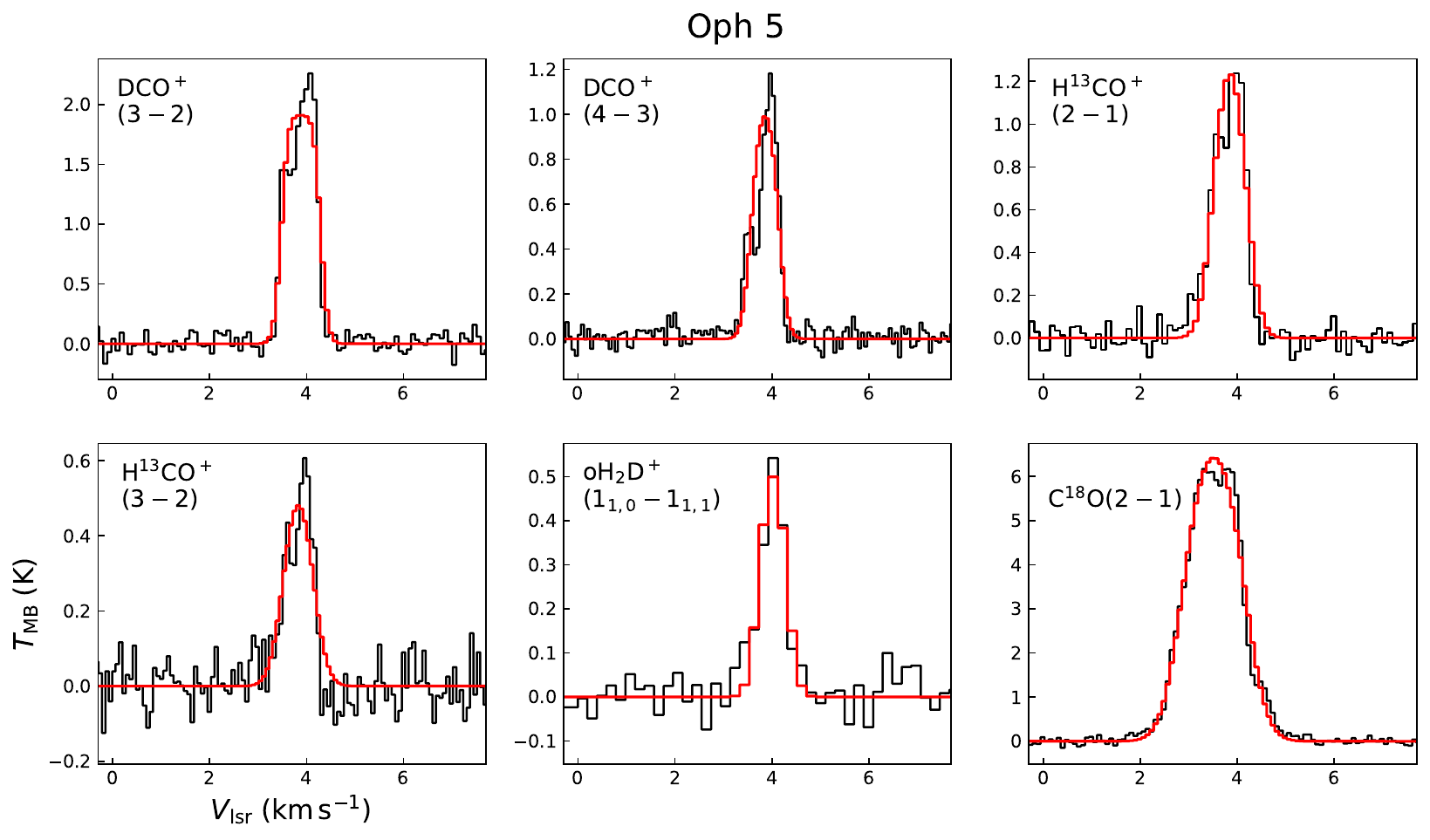}
    \caption{Same as Fig.~\ref{fig:first_app}, but for Oph 5.}
\end{figure}
\begin{figure}[!h]
    \centering
    \includegraphics[width=\linewidth]{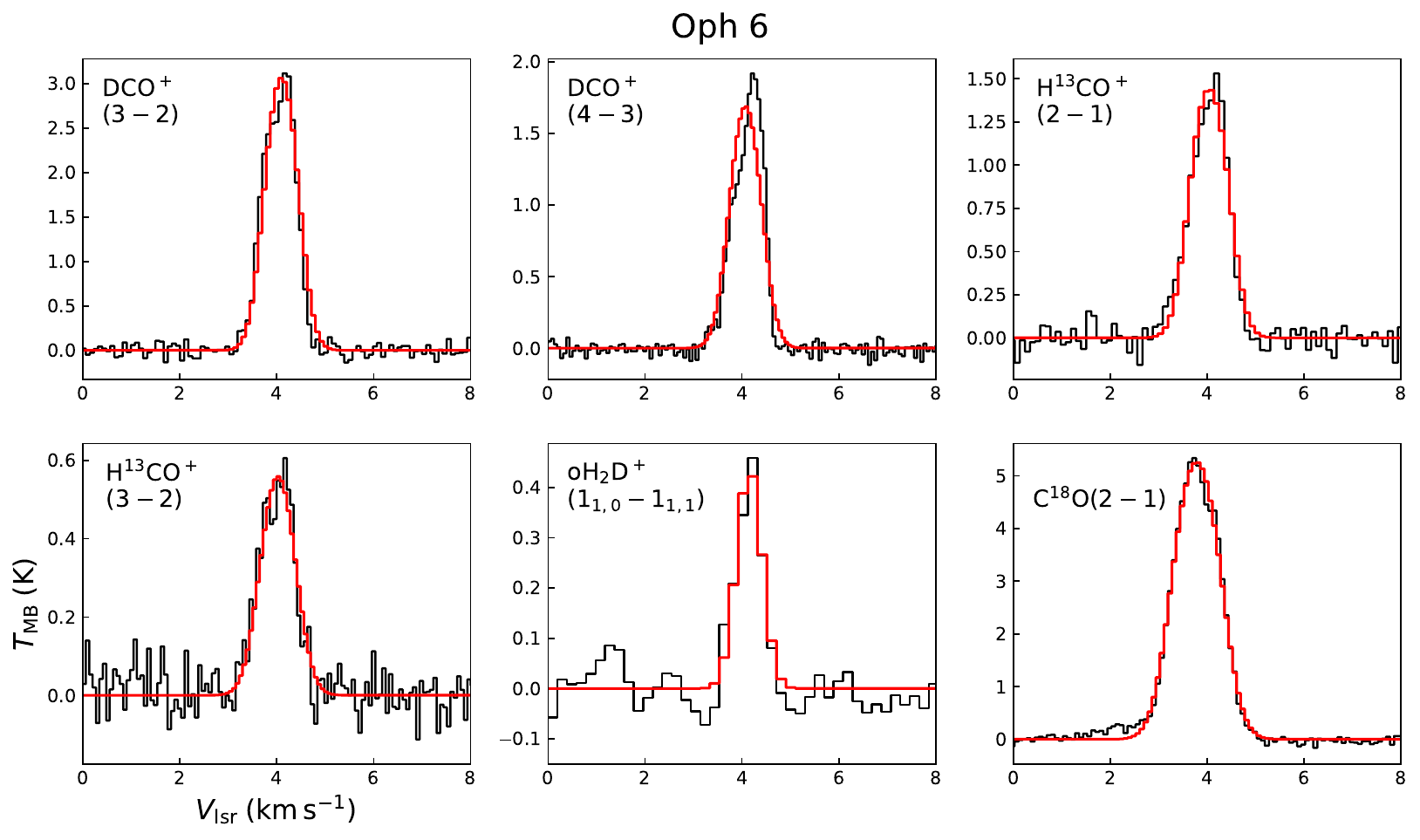}
    \caption{Same as Fig.~\ref{fig:first_app}, but for Oph 6.}
\end{figure}

\begin{figure}[!h]
    \centering
    \includegraphics[width=\linewidth]{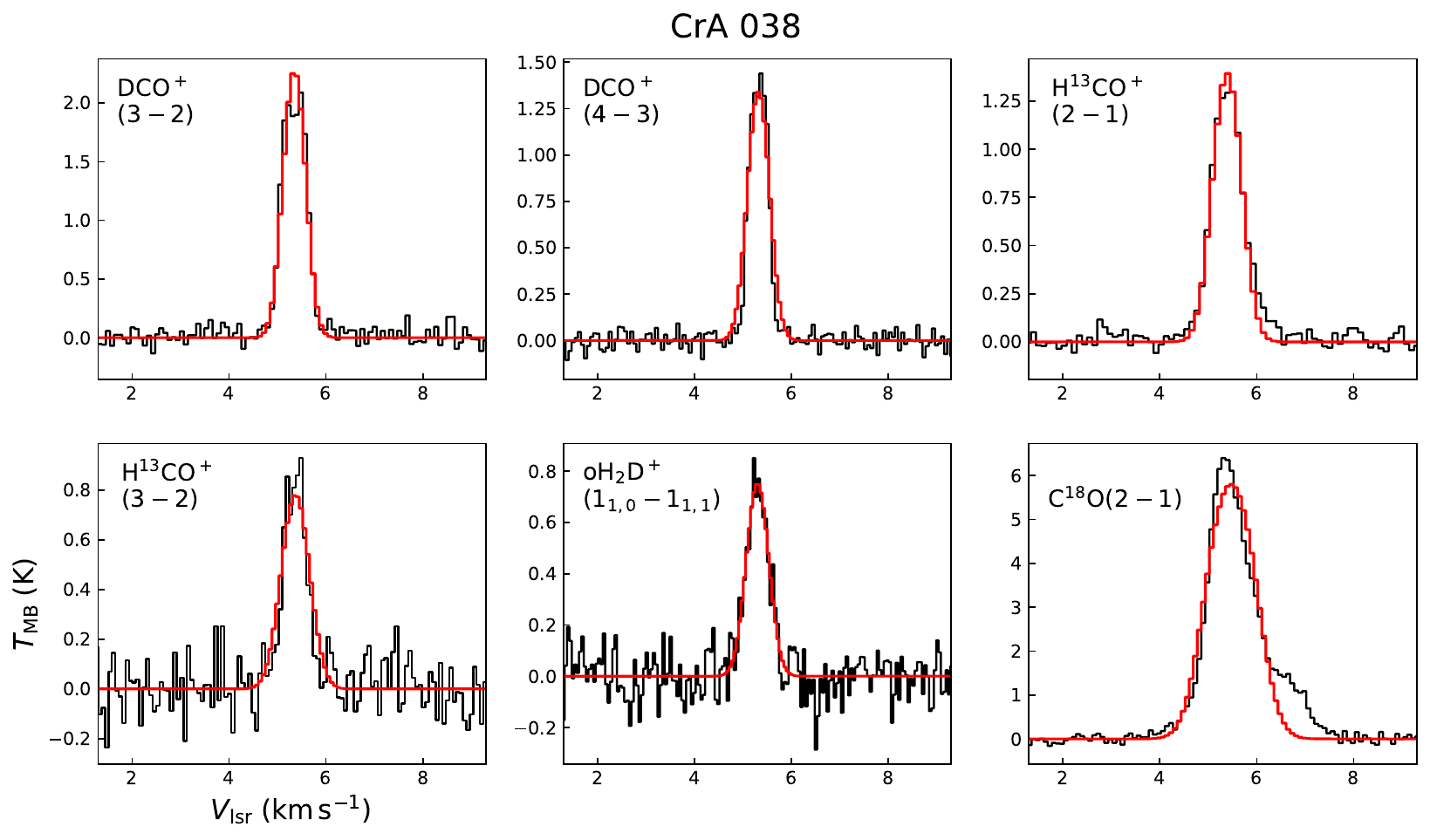}
    \caption{Same as Fig.~\ref{fig:first_app}, but for CrA 038.}
\end{figure}

\begin{figure}[!h]
    \centering
    \includegraphics[width=\linewidth]{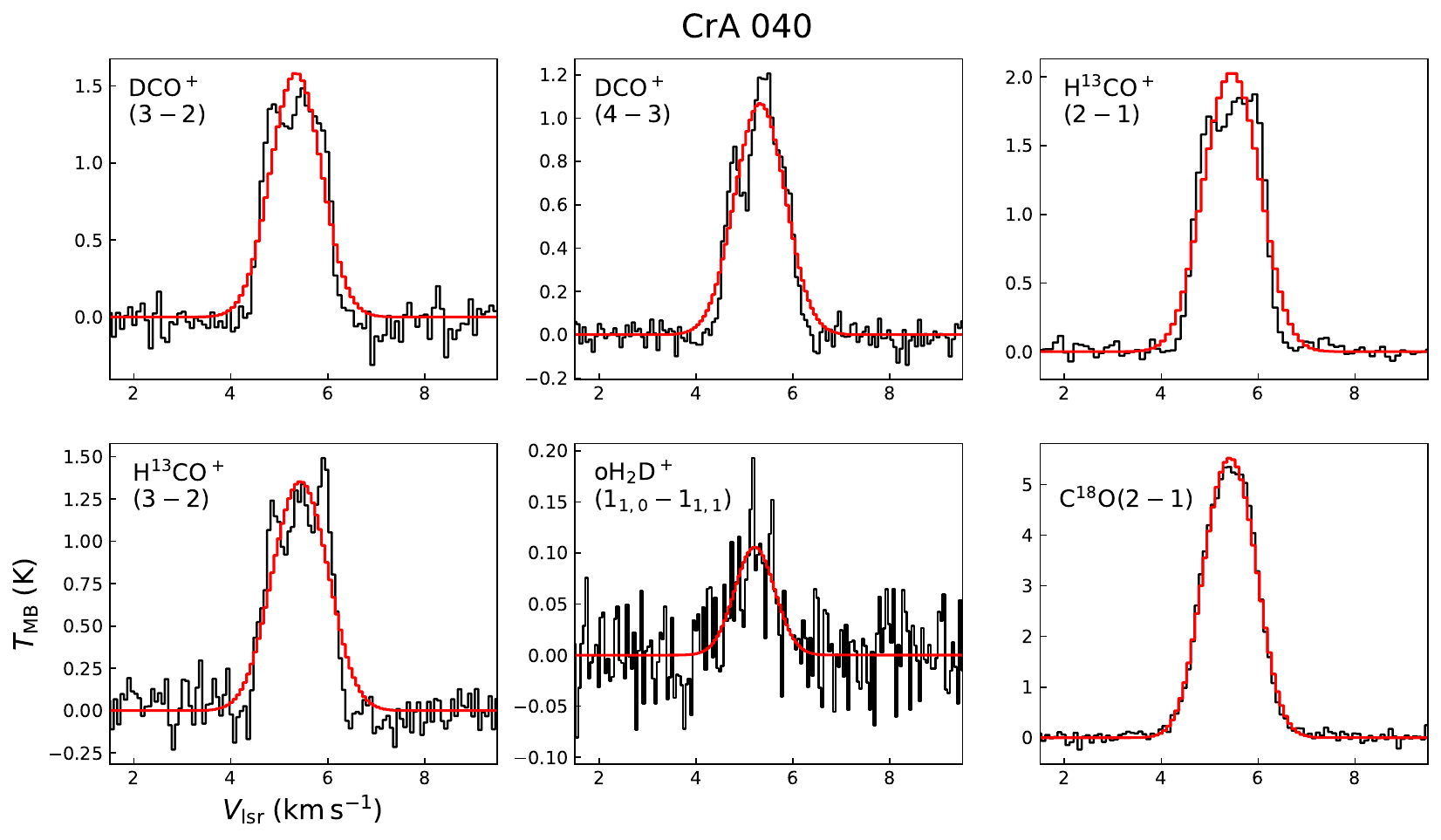}
    \caption{Same as Fig.~\ref{fig:first_app}, but for CrA 040.}
\end{figure}

\begin{figure}[!h]
    \centering
    \includegraphics[width=\linewidth]{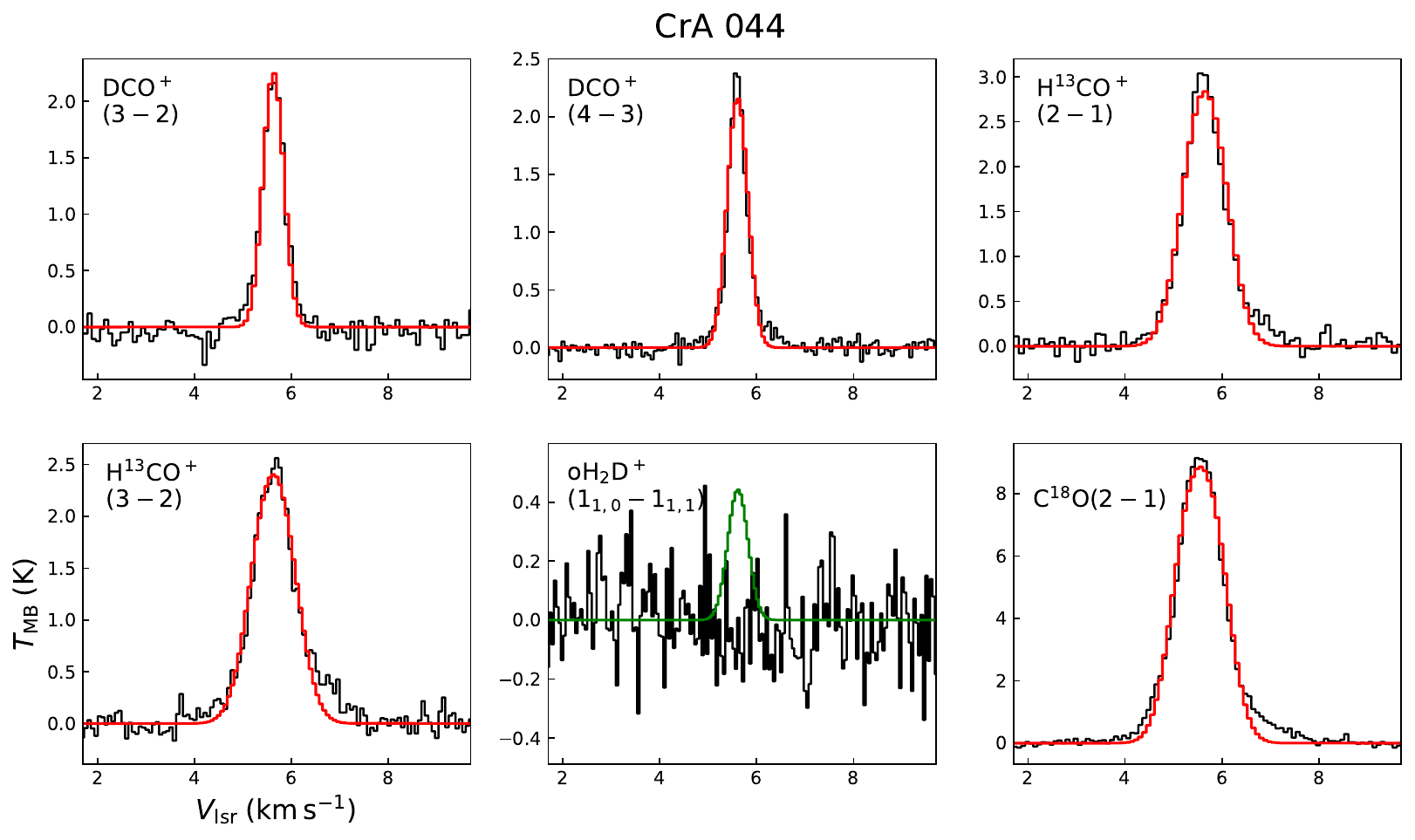}
    \caption{Same as Fig.~\ref{fig:first_app}, but for CrA 044. In this case, the \olineh line is not detected given the sensitivity. The green curve shows the $3\sigma$ upper limit obtained from the column density upper limit listed in Table~\ref{tab:col_dens}, and the centroid velocity and linewidth of the \dcop fit.}
\end{figure}
\begin{figure}[!h]
    \centering
    \includegraphics[width=\linewidth]{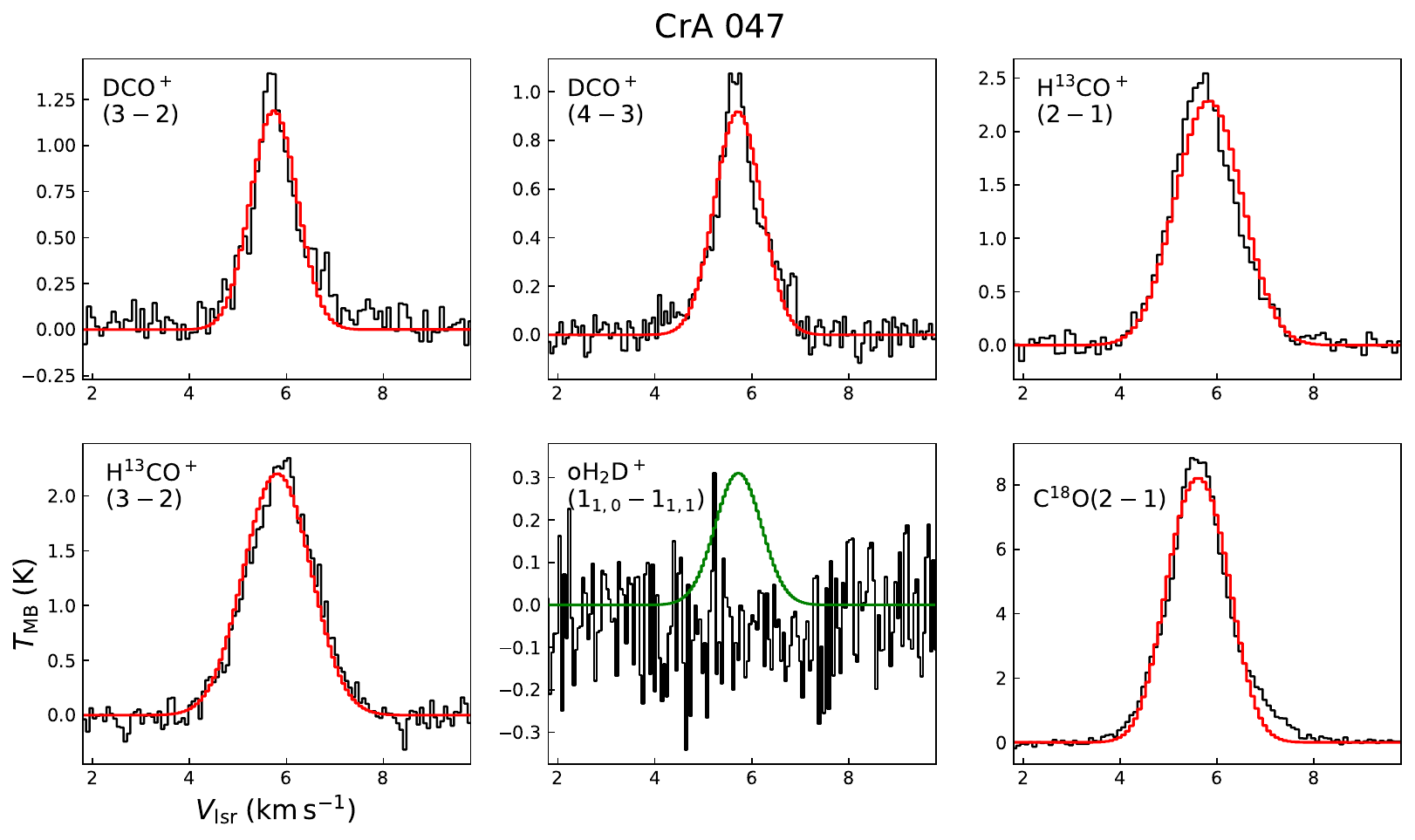}
    \caption{Same as Fig.~\ref{fig:first_app}, but for CrA 047. In this case, the \olineh line is not detected given the sensitivity. The green curve shows the $3\sigma$ upper limit obtained from the column density upper limit listed in Table~\ref{tab:col_dens}, and the centroid velocity and {velocity dispersion} of the \dcop fit.}
\end{figure}
\begin{figure}[!h]
    \centering
    \includegraphics[width=\linewidth]{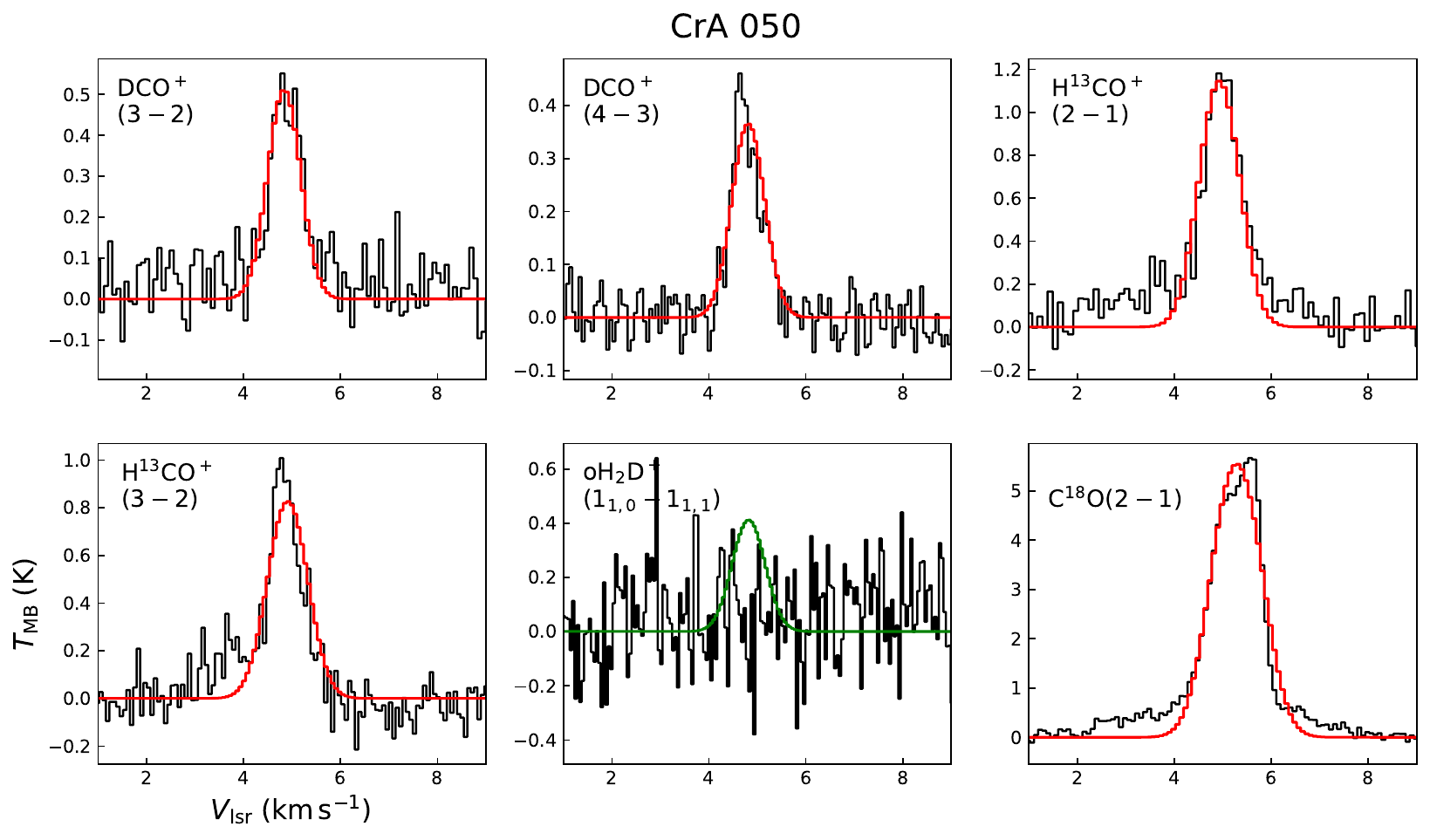}
    \caption{Same as Fig.~\ref{fig:first_app}, but for CrA 050. In this case, the \olineh line is not detected given the sensitivity. The green curve shows the $3\sigma$ upper limit obtained from the column density upper limit listed in Table~\ref{tab:col_dens}, and the centroid velocity and {velocity dispersion} of the \dcop fit.}
\end{figure}

\begin{figure}[!h]
    \centering
    \includegraphics[width=\linewidth]{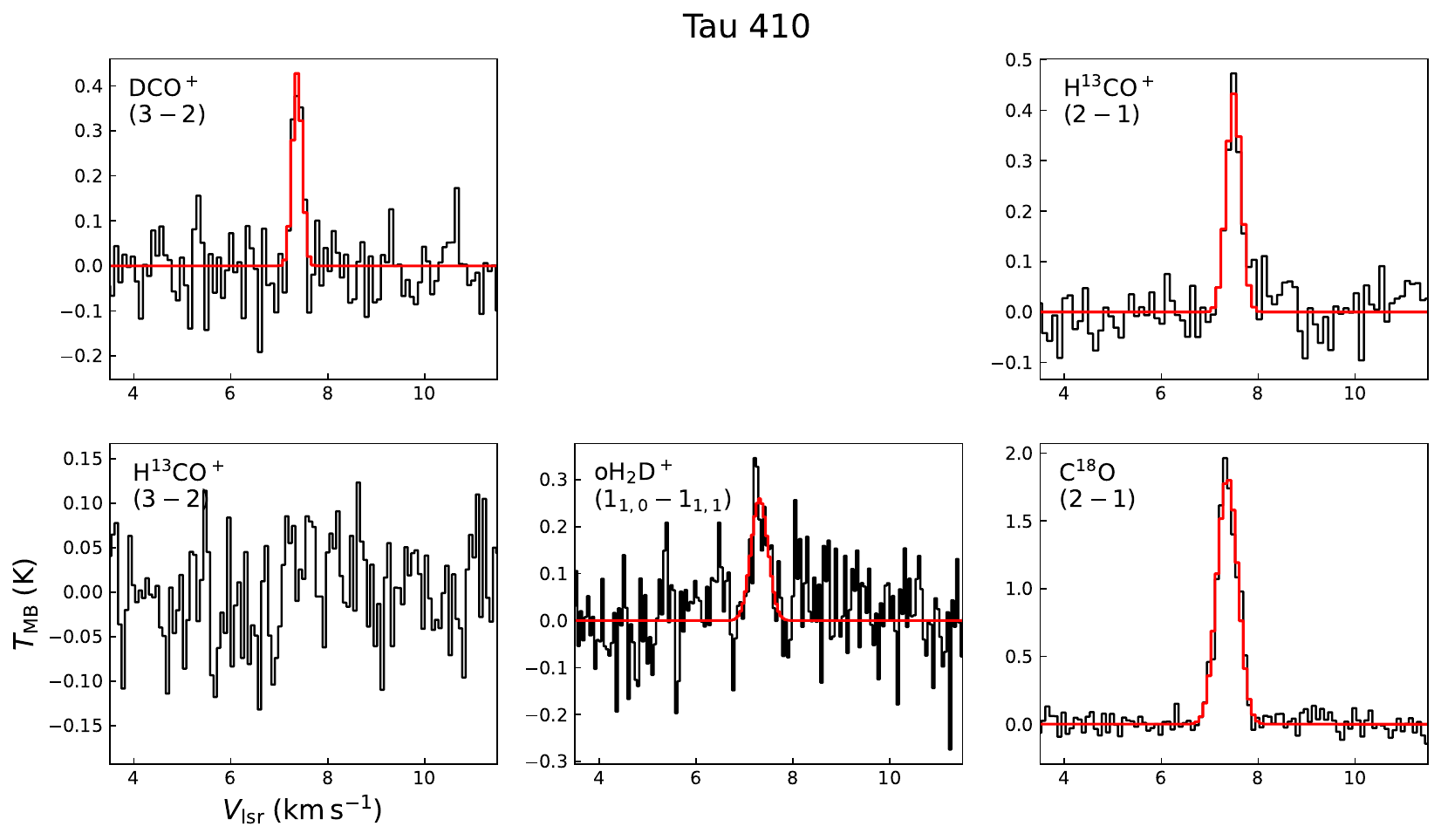}
    \caption{Same as Fig.~\ref{fig:first_app}, but for Tau 410.}
\end{figure}

\begin{figure}[!h]
    \centering
    \includegraphics[width=\linewidth]{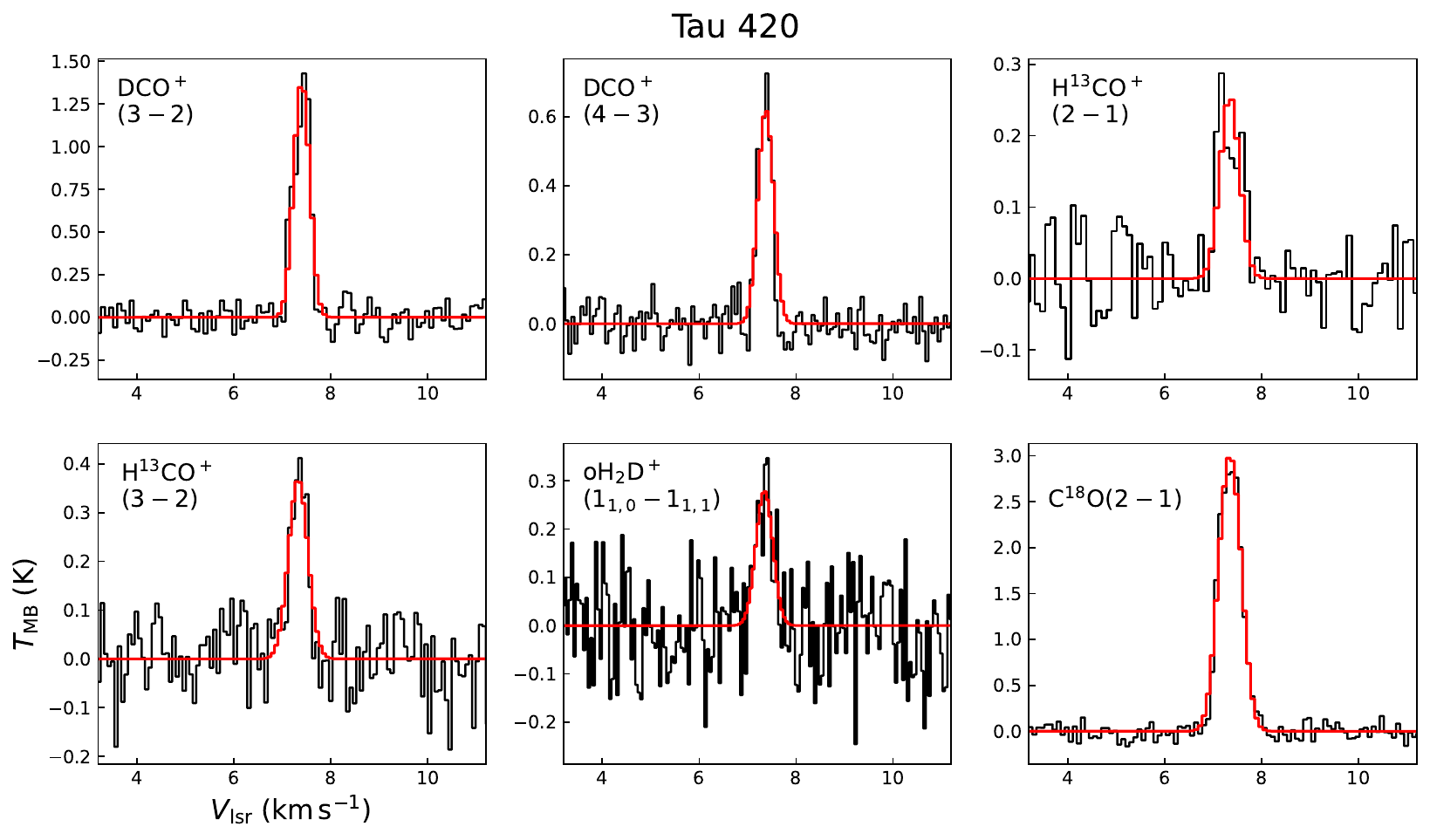}
    \caption{Same as Fig.~\ref{fig:first_app}, but for Tau 420.}
\end{figure}
\begin{figure}[!h]
    \centering
    \includegraphics[width=\linewidth]{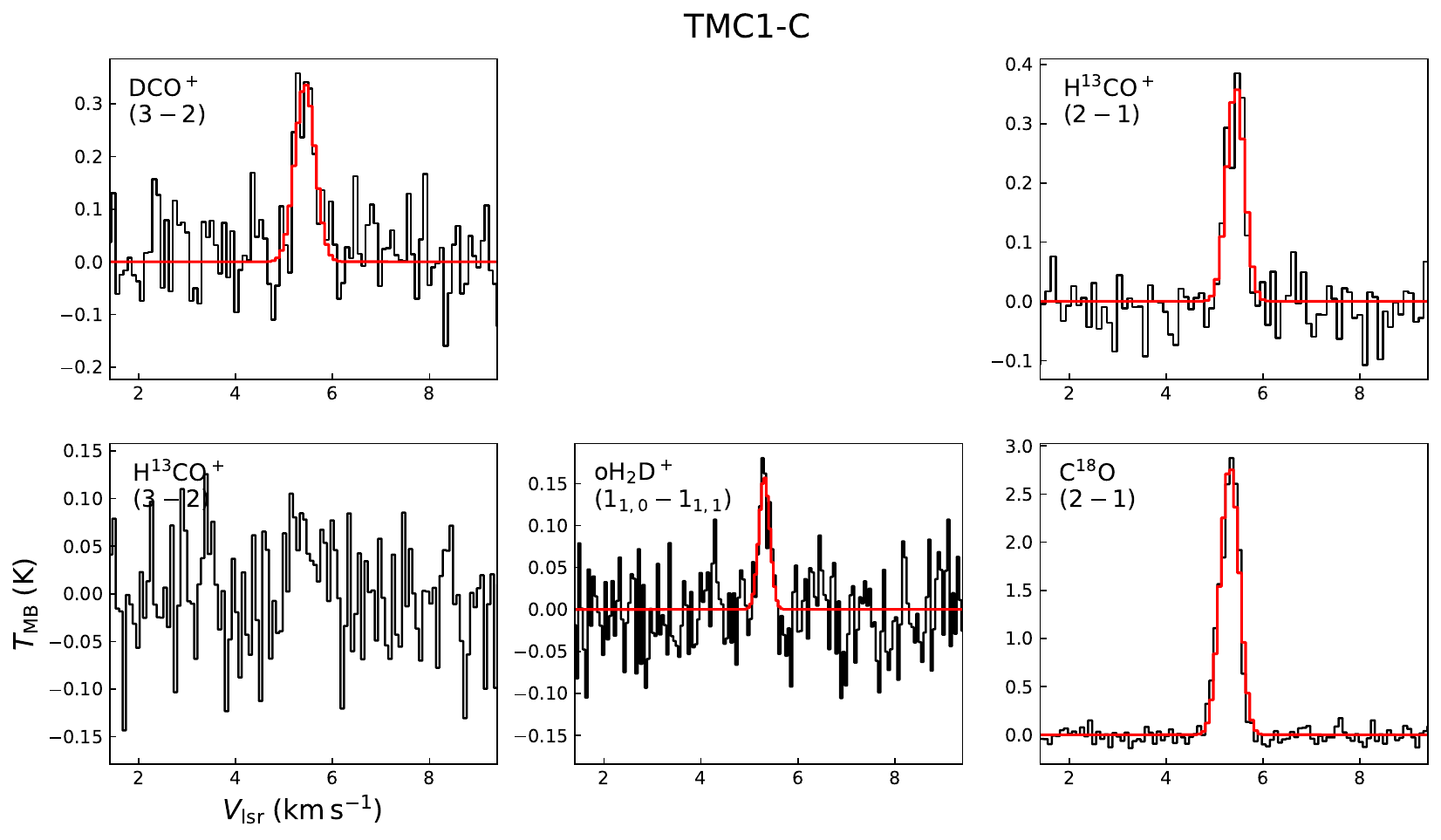}
    \caption{Same as Fig.~\ref{fig:first_app}, but for TMC1-C.}
\end{figure}
\begin{figure}[!h]
    \centering
    \includegraphics[width=\linewidth]{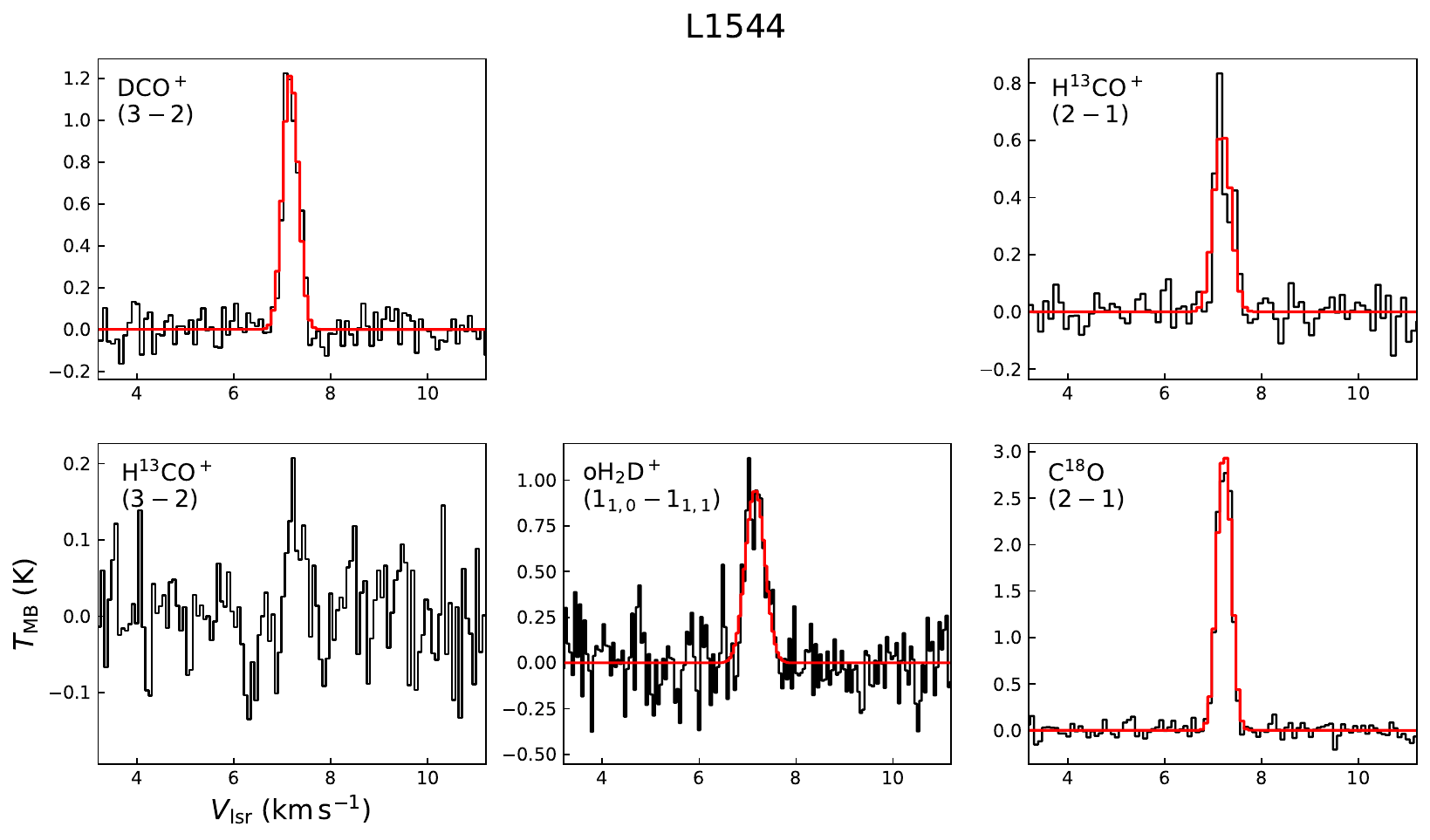}
    \caption{Same as Fig.~\ref{fig:first_app}, but for L1544.}
\end{figure}
\begin{figure}[!h]
    \centering
    \includegraphics[width=\linewidth]{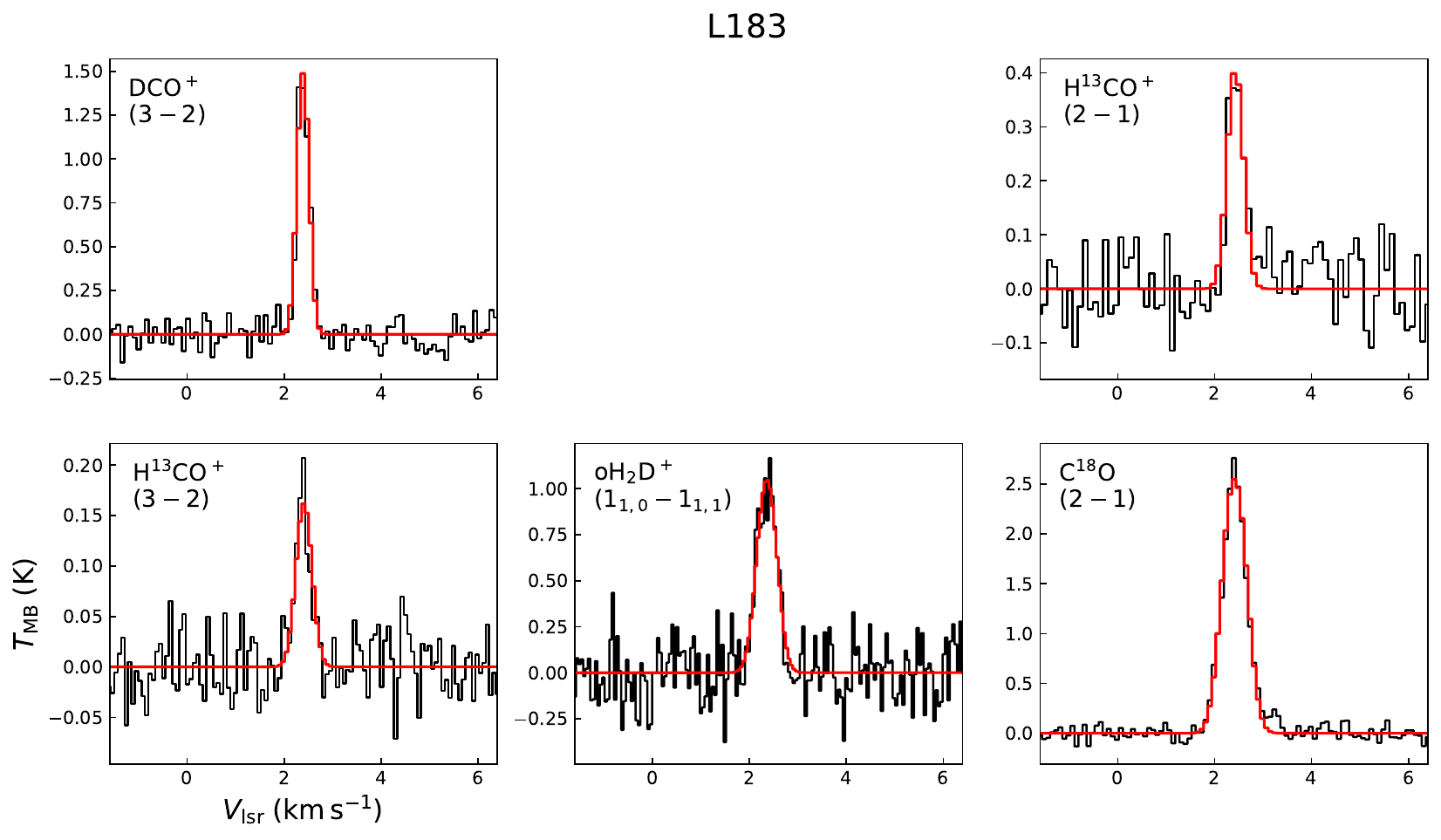}
    \caption{Same as Fig.~\ref{fig:first_app}, but for L183.}
\end{figure}
\begin{figure}[!h]
    \centering
    \includegraphics[width=\linewidth]{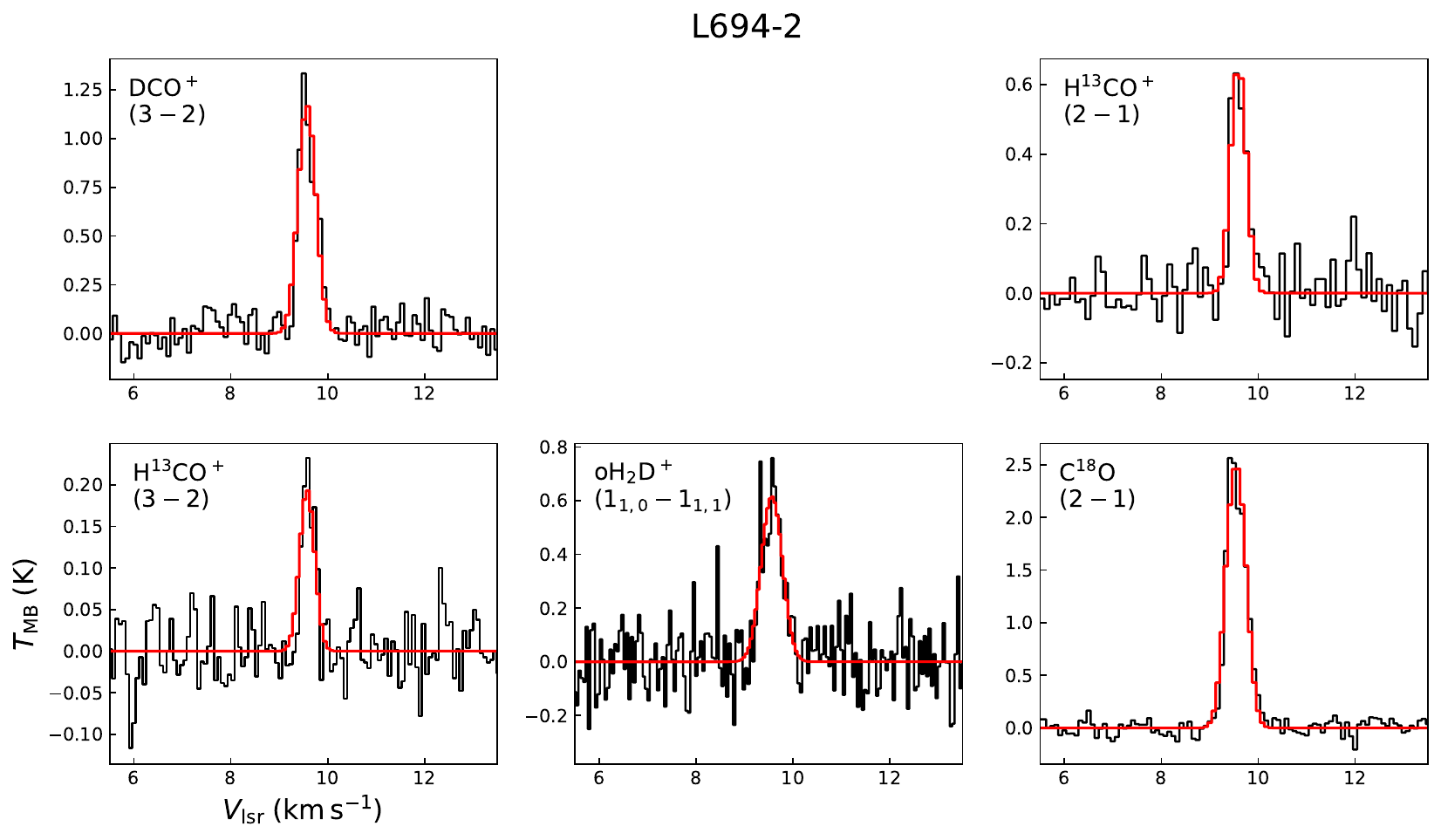}
    \caption{Same as Fig.~\ref{fig:first_app}, but for L694-2.}
\end{figure}

\begin{sidewaystable*}[!h]
\centering
    \renewcommand{\arraystretch}{1.2}
\tiny
\caption{The table summarises the best-fit values for the centroid velocity $V_\mathrm{lsr}$ and the {velocity dispersion} $\sigma\rm _V$ obtained with \textsc{pyspeckit}, together with the estimated peak opacity of each transition.}
\label{tab:fit_params}
\begin{tabular}{c|cccc|cccc|ccc|ccc}
\hline \hline
Source  &   \multicolumn{4}{c|}{\dcop} & \multicolumn{4}{c|}{\htcop} & \multicolumn{3}{c|}{\ohhdp} &\multicolumn{3}{c}{\cdo}\\
       & $V_\mathrm{lsr}$&$\sigma_V$ & $\tau$& $\tau$& $V_\mathrm{lsr}$&$\sigma_V$ & $\tau$& $\tau$& $V_\mathrm{lsr}$ &$\sigma_V$& $\tau $& $V_\mathrm{lsr}$& $\sigma_V$ & $\tau$  \\
        
      & \kms &  \kms &  $3-2$    & $4-3$  & \kms &  \kms &  $2-1$    &  $3-2$ & \kms &  \kms &   $1_{1,0}-1_{1,1}$   & \kms &  \kms &  $2-1$    \\
 \hline
Tau 410	&	$7.329\pm0.016$	&	$0.099\pm0.016$&	0.07	&		-	&	$7.442\pm0.014$	&	$0.151\pm0.014$&	0.12	&		-	&	$7.29 \pm0.03$	&	$0.17\pm0.03$  &	0.06	&	$7.317\pm0.006$	&	$0.204\pm0.005$&	0.31	\\
Tau 420	&	$7.357\pm0.005$	&	$0.157\pm0.005$&	0.32	&	0.17	&	$7.293\pm0.017$	&	$0.200\pm0.017$&	0.01	&	0.02	&	$7.32\pm0.03 $	&	$0.17\pm0.03$  &	0.09	&	$7.302\pm0.004$	&	$0.215\pm0.004$&	0.46	\\
TMC1-C	&	$5.39\pm0.03$	&	$0.21 \pm0.03 $&	0.06	&		-	&	$5.380\pm0.017$	&	$0.180\pm0.017$&	0.10	&		-	&	$5.29\pm0.02 $	&	$0.11\pm0.02$  &	0.04	&	$5.267\pm0.004$	&	$0.180\pm0.003$&	0.53	\\
L1544	&	$7.127\pm0.007$	&	$0.154\pm0.006$&	0.26	&		-	&	$7.136\pm0.012$	&	$0.172\pm0.012$&	0.23	&		-	&	$7.129\pm0.017$	&	$0.191\pm0.017$&	0.34	&	$7.182\pm0.003$	&	$0.135\pm0.002$&	0.74	\\
L183	&	$2.347\pm0.005$	&	$0.119\pm0.005$&	0.39	&		-	&	$2.374\pm0.013$	&	$0.165\pm0.013$&	0.14	&	0.08	&	$2.335\pm0.015$	&	$0.193\pm0.014$&	0.45	&	$2.363\pm0.004$	&	$0.227\pm0.003$&	0.69	\\
L429	&	$6.628\pm0.009$	&	$0.226\pm0.008$&	0.34	&		-	&	$6.697\pm0.012$	&	$0.245\pm0.012$&	0.32	&	0.12	&	$6.644\pm0.013$	&	$0.236\pm0.012$&	0.39	&	$6.864\pm0.006$	&	$0.192\pm0.005$&	0.52	\\
L694-2	&	$9.530\pm0.008$	&	$0.177\pm0.008$&	0.28	&		-	&	$9.538\pm0.011$	&	$0.153\pm0.010$&	0.38	&	0.15	&	$9.54 \pm0.02$	&	$0.211\pm0.019$&	0.23	&	$9.493\pm0.004$	&	$0.187\pm0.004$&	0.63	\\
Oph 1	&	$4.424\pm0.005$	&	$0.164\pm0.005$&	0.71	&	0.23	&	$4.450\pm0.010$	&	$0.158\pm0.010$&	0.36	&	0.12	&	$4.45\pm0.02$	&	$0.27 \pm0.02 $&	0.49	&	$4.575\pm0.009$	&	$0.370\pm0.008$&	0.39	\\
Oph 2	&	$4.485\pm0.004$	&	$0.242\pm0.004$&	0.59	&	0.33	&	$4.488\pm0.008$	&	$0.227\pm0.008$&	0.58	&	0.26	&	$4.473\pm0.018$	&	$0.213\pm0.018$&	0.14	&	$4.425\pm0.002$	&	$0.229\pm0.002$&	0.93	\\
Oph 3-1\tablefootmark{a}	&	$3.59\pm0.02$	&	$0.267\pm0.012$&	0.38	&	0.27	&	$3.45\pm0.02 $	&	$0.220\pm0.014$&	0.60	&	0.28	&	$3.55\pm0.18$	&	$0.22\pm0.13 $ &	0.02	&	$3.318\pm0.009$	&	$0.269\pm0.005$&	0.91	\\
Oph 3-2\tablefootmark{a}	&	$4.166\pm0.016$	&	$0.251\pm0.008$&	0.25	&	0.24	&	$4.13\pm0.02 $	&	$0.314\pm0.013$&	0.69	&	0.41	&	$4.11\pm0.09 $	&	$0.23 \pm0.07 $&	0.02	&	$4.155\pm0.012$	&	$0.306\pm0.009$&	0.64	\\
Oph 4-1\tablefootmark{a}	&	$3.231\pm0.007$	&	$0.183\pm0.007$&	0.80	&	0.26	&	$3.244\pm0.010$	&	$0.210\pm0.010$&	0.75	&	0.34	&	$3.35\pm0.18 $	&	$0.28\pm0.15  $&	0.15	&	$2.966\pm0.012$	&	$0.257\pm0.004$&	0.66	\\
Oph 4-2\tablefootmark{a}	&	$3.885\pm0.003$	&	$0.153\pm0.004$&	0.64	&	0.33	&	$3.872\pm0.010$	&	$0.171\pm0.009$&	0.60	&	0.28	&	$3.94\pm0.04 $	&	$0.19 \pm0.03 $&	0.18	&	$3.68 \pm0.02 $	&	$0.391\pm0.011$&	0.93	\\
Oph 5	&	$3.842\pm0.005$	&	$0.203\pm0.004$&	5.01	&	1.26	&	$3.795\pm0.008$	&	$0.288\pm0.008$&	0.65	&	0.30	&	$3.92 \pm0.02$	&	$0.229\pm0.018$&	0.83	&	$3.468\pm0.004$	&	$0.467\pm0.004$&	1.22	\\
Oph 6	&	$4.056\pm0.004$	&	$0.302\pm0.004$&	0.77	&	0.43	&	$3.991\pm0.007$	&	$0.337\pm0.008$&	0.93	&	0.41	&	$4.06\pm0.02$	&	$0.26 \pm0.02 $&	0.12	&	$3.736\pm0.003$	&	$0.417\pm0.003$&	0.88	\\
Oph D	&	$3.472\pm0.002$	&	$0.140\pm0.002$&	0.75	&	0.42	&	$3.513\pm0.005$	&	$0.139\pm0.006$&	1.03	&	0.41	&	$3.387\pm0.018$	&	$0.191\pm0.017$&	0.23	&	$3.520\pm0.004$	&	$0.210\pm0.004$&	0.53	\\
CrA 038	&	$5.305\pm0.004$	&	$0.218\pm0.004$&	0.39	&	0.25	&	$5.336\pm0.009$	&	$0.282\pm0.008$&	0.38	&	0.26	&	$5.278\pm0.014$	&	$0.227\pm0.013$&	0.17	&	$5.425\pm0.008$	&	$0.429\pm0.007$&	1.06	\\
CrA 040	&	$5.304\pm0.009$	&	$0.506\pm0.009$&	0.19	&	0.15	&	$5.401\pm0.010$	&	$0.540\pm0.009$&	0.45	&	0.36	&	$5.19\pm0.05 $	&	$0.41 \pm0.05 $&	0.02	&	$5.370\pm0.003$	&	$0.478\pm0.003$&	0.73	\\
CrA 044	&	$5.590\pm0.003$	&	$0.209\pm0.003$&	0.14	&	0.15	&	$5.596\pm0.005$	&	$0.419\pm0.005$&	0.42	&	0.42	&				-	&		-	&		-			&	$5.505\pm0.004$	&	$0.422\pm0.003$&	1.10	\\
CrA 047	&	$5.698\pm0.009$	&	$0.472\pm0.009$&	0.11	&	0.09	&	$5.775\pm0.008$	&	$0.658\pm0.008$&	0.25	&	0.28	&				-	&		-	&		-			&	$5.562\pm0.006$	&	$0.548\pm0.005$&	0.89	\\
CrA 050	&	$4.795\pm0.016$	&	$0.338\pm0.016$&	0.05	&	0.04	&	$4.876\pm0.013$	&	$0.404\pm0.012$&	0.19	&	0.17	&				-	&		-	&		-			&	$5.241\pm0.007$	&	$0.476\pm0.007$&	0.56	\\
CrA 151	&	$5.603\pm0.004$	&	$0.215\pm0.003$&	0.34	&	0.23	&	$5.564\pm0.009$	&	$0.173\pm0.009$&	0.54	&	0.26	&	$5.62\pm0.03$	&	$0.20\pm0.03$  &	0.14	&	$5.456\pm0.006$	&	$0.298\pm0.005$&	0.67	\\
\hline
\end{tabular}
\tablefoot{
\tablefoottext{a}{For Oph 3 and Oph 4 we report the fit values of both velocity components.} 
}
\end{sidewaystable*}

\clearpage

\section{LVG modelling of the \htcop lines\label{app:RADEX}}

Given the availability of multiple transitions of \htcop for almost all the targets, we have used these data to perform simple LVG (large-velocity-gradient) calculation using the \textsc{radex} software (\citealt{vanderTak07}; Dwiriyanti et al., subm.). This allows us to \textit{i)} infer $n\rm (H_2)$ independently from the \textit{Herschel} data; \textit{ii)} re-evaluate the path length $L = N\mathrm{(H_2)}/n\rm (H_2)$; and \textit{iii)} confirm the sub-thermal excitation for this species. {We have used \textsc{radex} to compute escape probabilities, adopting the LVG approach.} The {LAMDA (Leiden Atomic and Molecular DAtabase)} file for \htcop has been taken from the EMAA database\footnote{\url{https://emaa.osug.fr}.}. We consider para-$\rm H_2$ the main collisional partner. For the sake of simplicity, we have excluded from this analysis the two sources with clear multiple velocity components in all lines (Oph 3 and Oph 4), and the 3 sources with only upper limits for the \htcop (3-2) transition (Tau 410, TMC1-C, L1544). The number of analysed targets is 15. \par
We proceeded to input into the RADEX calculation the \htcop column density and the line width (FWHM) derived in Sect.~\ref{sec:Ncols}. We used the dust temperatures in Table~\ref{tab:cores} as a proxy for the gas temperature, and the gas volume densities obtained from the $N\mathrm{(H_2)}$ values (Table~\ref{tab:cores}, assuming $L=0.1 \rm \,pc$). We refer to this quantity as $n\mathrm{(H_2)}_{\rm cont}$, as it is derived from the \textit{Herschel} continuum data. \par
\begin{figure*}
    \centering
    \includegraphics[width=\linewidth]{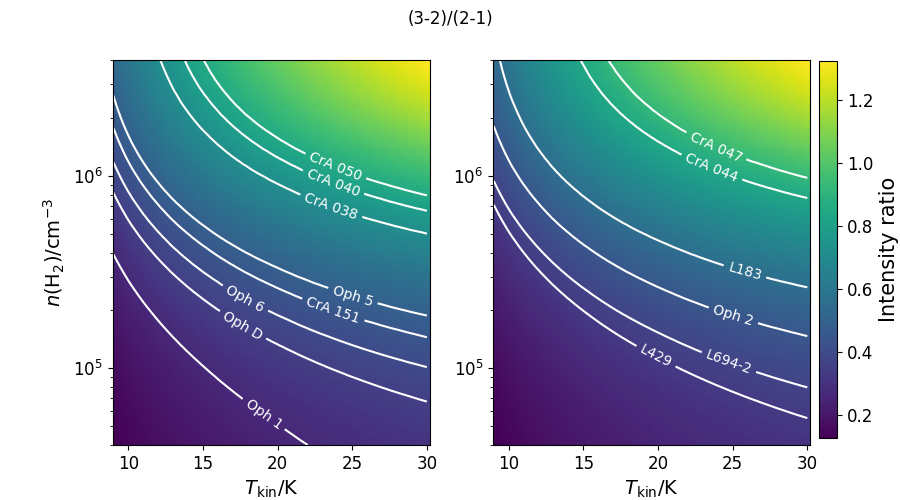}
    \caption{ \htcop line intensity ratio (3-2)/(2-1) as a function of H${_2}$ volume density versus gas kinetic temperature ($T_{\rm kin}$). The plot has been obtained with \textsc{radex}, assuming the average \htcop column density measured in the sample ($9.4\times 10^{11} \, \rm cm^{-2}$). The curves, labelled by object, show the observed values.}
    \label{fig:radex_results}
\end{figure*}

We checked the results, comparing the modelled line intensities with the observed ones. {Table~\ref{tab:radex} reports the $n\mathrm{(H_2)}_{\rm cont}$ values for each source, together with the lines peak intensity, both observed and modelled by \textsc{radex}. We also list the ratio $T  _{\rm peak} ^{\rm mod}/T \rm _{peak} ^{obs}$. These} results show that for 14 out of 15 targets, the (2-1) line intensity is reproduced within a factor of 2, and for the (3-2) lines, this holds for 6 targets. Since the (3-2) line has a better angular resolution and higher critical density, this suggests that the $n\mathrm{(H_2)}_{\rm cont}$ values underestimate the central volume density. This is expected due to the fact that this quantity is computed from the Herschel low-resolution data, and it represents an average along the line-of-sight and on the beam. {Figure~\ref{fig:radex_results} shows the \textsc{radex} results concerning the (3-2) to (2-1) line intensity ratio for a grid of gas temperature and $n\rm (H_2)$ values. We overplot the curves corresponding to the observed cores. The plot shows that the \htcop observations are consistent with volume density values from several $\times 10^4 \, \rm cm^{-3}$ to $10^6 \, \rm cm^{-3}$, given the typical cores' temperatures of $10-20\, \rm K$. } \par
In order to quantify the underestimation, we have increased $n\mathrm{(H_2)}$ and we ran the LVG computation again. First, we input $n\mathrm{(H_2)} = 2 \times n\mathrm{(H_2)}_{\rm cont}$. {This is sufficient to reproduce the line instensities within a factor of 2 in all sources except two}. The only targets where the emission is still underestimated are L183 (10\% for the 2-1, 60\% for the 3-2), for which a correction $n\mathrm{(H_2)} = 3 \times n\mathrm{(H_2)}_{\rm cont}$ is needed, and Tau 420, which is a peculiar source due to the fact that the (3-2) line is brighter than the (2-1) (this is also the only source where we see suprathermal excitation in \htcop, see Table~\ref{tab:col_dens} and Fig.~\ref{fig:hcop_dcop}). We can conclude that the $n\mathrm{(H_2)}_{\rm cont}$ volume densities are underestimated by a factor of $\sim 2$. \par

\begin{figure}
    \centering
    \includegraphics[width=\linewidth]{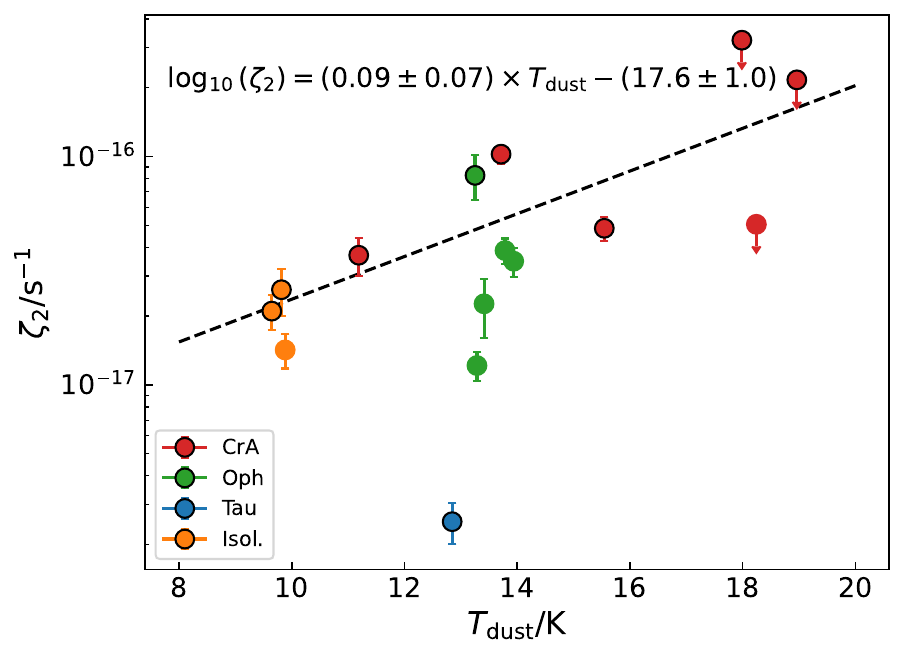}
    \caption{Plot of \crir as a function of \tdust, where the $n\rm (H_2)$ values have been re-evaluated from the \htcop line emission, as explained in the Text. The sources where we increased $n\mathrm{(H_2)}_{\rm cont}$ in the LVG calculations are shown with black edges. The dashed black curve shows the linear best fit to the data, in log-linear space.}
    \label{fig:CRIR_newnh2}
\end{figure}

We have re-computed the \crir values assuming $L = 0.05\rm \, pc$ for those 9 sources that needs the correction. The results are shown in Fig~\ref{fig:CRIR_newnh2}, as a function of the dust temperature. We still detect the increase of \crir with increasing \tdust. The black curve shows the linear best fit to the data. The trend is shallower and the uncertainties are higher (due to the reduced statistics), but our results still hold. 
We highlight that the LVG methodology entrails a number of uncertainties, the main ones being \textit{i)} the $T_\mathrm{gas} = T\rm _{dust}$ assumption; \textit{ii)} the source geometry, which is ignored. \par
The \textsc{radex} analysis confirms that the \htcop transition are subthermally excited. The $\tau$ values computed by \textsc{radex} are in general in good agreement with those listed in Table~\ref{tab:fit_params}, confirming that the lines are optically thin.

\begin{table}[!h]

\caption{Results of the \textsc{radex} analysis. For each source, we report the $\rm H_2$ volume density estimated from \textit{Herschel}, the modelled vs observed peak temperature of the \htcop lines, and their ratio $T  _{\rm peak} ^{\rm mod}/T \rm _{peak} ^{obs}$.  \label{tab:radex}}
\begin{tabular}{c|c|ccc|ccc}
\hline
\hline
        &        & \multicolumn{3}{c|}{(2-1)}& \multicolumn{3}{|c}{(3-2)} \\
Source &  $n\mathrm{(H_2)}_{\rm cont}$   &  $T \rm _{peak} ^{obs}$  &  $T \rm _{peak} ^{mod} $ & ratio &$T \rm _{peak} ^{obs} $  & $T \rm _{peak} ^{mod} $   &  ratio       \\
      &      \multicolumn{1}{c|}{$10^5\,\rm cm^{-3}$} & K & K & \multicolumn{1}{c|}{}&K &K &\\
\hline
 Tau 420 & 1.3 & 0.3 & 0.1 &  0.4 &  0.4 &  0.03 &  0.1 \\
 L183    & 1.9 & 0.4 & 0.2 &  0.6 &  0.2 &  0.04 &  0.2 \\
 L429    & 1.7 & 0.5 & 0.4 &  0.8 &  0.2 &  0.1  &  0.5 \\
 L694-2  & 1.4 & 0.6 & 0.4 &  0.7 &  0.2 &  0.1  &  0.3 \\
 Oph 1   & 1.2 & 0.5 & 0.5 &  0.9 &  0.1 &  0.1  &  0.8 \\
 Oph 2   & 1.7 & 1.0 & 1.0 &  1.1 &  0.4 &  0.3  &  0.6 \\
 Oph 5   & 1.7 & 1.2 & 1.2 &  1.0 &  0.6 &  0.3  &  0.6 \\
 Oph 6   & 1.5 & 1.5 & 1.5 &  1.0 &  0.6 &  0.4  &  0.7 \\
 Oph D   & 0.5 & 1.4 & 0.8 &  0.6 &  0.5 &  0.2  &  0.3 \\
 CrA 038 & 2.0 & 1.3 & 1.0 &  0.8 &  0.9 &  0.3  &  0.3 \\
 CrA 040 & 1.5 & 1.9 & 1.2 &  0.7 &  1.5 &  0.4  &  0.3 \\
 CrA 044 & 3.4 & 3.0 & 2.6 &  0.8 &  2.6 &  1.2  &  0.5 \\
 CrA 047 & 3.2 & 2.5 & 1.9 &  0.8 &  2.3 &  0.9  &  0.4 \\
 CrA 050 & 1.1 & 1.2 & 0.6 &  0.5 &  1.0 &  0.2  &  0.2 \\
 CrA 151 & 1.3 & 1.2 & 0.7 &  0.6 &  0.5 &  0.2  &  0.3 \\
\hline
\end{tabular}
\end{table}
\end{document}